\documentclass[
  twocolumn,
  reprint,
  nofootinbib,
  amsmath,amssymb,
  pra,
  10pt
]{revtex4-2}

\usepackage{graphicx,color}
\usepackage{bm}
\usepackage{braket}
\usepackage{ulem}
\usepackage{hyperref}
\usepackage{adjustbox}
\usepackage{booktabs}

\definecolor{battleshipgrey}{rgb}{0.52, 0.52, 0.51}
\definecolor{cadet}{rgb}{0.33, 0.41, 0.47}
\definecolor{charcoal}{rgb}{0.21, 0.27, 0.31}

\hypersetup{
  colorlinks   = true,
  urlcolor     = battleshipgrey,
  linkcolor    = cadet,
  citecolor    = charcoal
}

\begin{document}

\preprint{APS/123-QED}
\title{Unravelling inter-channel quantum interference in below-threshold nonsequential double ionization with statistical measures}
\author{S. Hashim}
\author{C. Figueira de Morisson Faria}%
\email{c.faria@ucl.ac.uk}
\affiliation{%
Department of Physics \& Astronomy, University College London \\Gower Street London  WC1E 6BT, United Kingdom
}%

\date{\today}

\begin{abstract}
We present a systematic study of interchannel quantum interference in laser-induced nonsequential double ionization (NSDI) within the strong-field approximation. Focusing on the below-threshold intensity regime where the recollision–excitation with subsequent ionization (RESI) pathway dominates, we derive analytical phase conditions governing interference between distinct excitation channels for arbitrary driving fields. To quantify the interplay between channels resulting from a vast number of interfering processes, we introduce statistical metrics based on the Earth Mover’s Distance, allowing us to assess the relative weight of each channel’s contribution to the two-electron photoelectron momentum distributions (PMDs). We identify key factors that determine whether interchannel interference is appreciable such as comparable channel intensities, strong spatial overlap between the excited-state wavefunctions and the energy difference between contributing channels. We demonstrate that for linearly polarized few-cycle pulses, the typical intrachannel interference features associated with exchange, temporal shifts and combined exchange-temporal interference are retained with interchannel interference. Our findings establish a hierarchy of interference mechanisms in RESI and may provide practical guidelines for enhancing or suppressing interference in different regions of the momentum plane. The toolkit presented in this work is transferable to a wide range of interferometric schemes involving different excitation channels.

\end{abstract}

\maketitle

\section{Introduction}
\label{sec:intro}
Quantum interference has commanded significant attention in strong-field and attosecond physics in recent years, both for its role in applications like attosecond imaging of matter and electron–hole dynamics \cite{Krausz2009, Haessler2010, Marangos2016} and for their potential in quantum technologies \cite{Stammer2024}. 
In particular, whenever several quantum pathways contribute coherently to a given observable, phase differences between them may leave measurable signatures in the resulting spectra. Such situations arise naturally throughout quantum mechanics, even in foundational concepts such as the double-slit experiment. They are ubiquitous not only 
in strong-field ionization, but also in other research fields such as high-precision spectroscopy \cite{Haensch2006,Hall2006} and quantum-optical interferometry \cite{Caves1981,Giovannetti2004}, where the quantitative comparison of probability distributions forms an essential diagnostic tool. Techniques such as ultrafast holography \cite{Huismans2012}, phase-of-the-phase spectroscopy \cite{Bauer2015}, and chiral light–matter interactions \cite{Ayuso2018} exemplify how quantum interference can be useful for precise temporal and spatial control in ultrafast phenomena.

Most of these studies, however, have been performed considering a single-active electron, while investigations that identify and exploit quantum interference in strongly correlated two-electron systems are comparatively rare \cite{Hao2014, Maxwell2015, Maxwell2016, Quan2017}. 
Non-sequential double ionization (NSDI) is particularly suitable  for exploring such effects, as it combines electron–electron correlation with intense-field dynamics  (for reviews see \cite{Faria2011,Ho2012}). Thereby, an electron rescatters inelastically with its parent ion, transferring enough energy to the core to release a second electron \cite{Corkum1993, Becker2012}. In the low-frequency regime, NSDI may occur in two main ways: Electron Impact Ionization (EI)  and Recollision-Excitation wtih Subsequent Ionization (RESI). EI dominates when the recolliding electron transfers sufficient kinetic energy to the core so that the second electron can overcome the ionization potential of the singly ionized target. It is then released immediately, with quantum signatures largely suppressed by averaging over transverse momenta \cite{Fu2012}. However, RESI arises when recollision only excites a bound electron which then ionizes after a time delay, enabling observable interference effects between the photoelectron even after focal averaging \cite{Maxwell2016,Hao2014}. 

Within RESI, quantum interference can be characterized as intrachannel interference, stemming from multiple events within a single excitation pathway, and interchannel interference arising from interference between pathways through different excited states. 
The investigations in \cite{Maxwell2016,Hao2014}, performed for long pulses that can be approximated by monochromatic waves, show that interchannel interference can reshape the RESI distributions. In \cite{Hao2014}, this interference has been employed to break a fourfold symmetry present in incoherent RESI distributions, and appropriate coherent superpositions of excitation channels have been used in \cite{Maxwell2016} to obtain correlated,
anticorrelated, cross-, or ring-shaped RESI distributions.

Excitation to multiple intermediate channels underlies many interferometric schemes such as RABITT, high-harmonic interferometry, and multiphoton ionization studies \cite{Dudovich2006, Shafir2012}. However, the systematic investigation of interchannel interference conditions in RESI under arbitrary driving fields remains incomplete. Previous works addressed the general conditions for intrachannel interference, and applied these to few-cycle pules \cite{Hashim2024} and bichromatic fields \cite{Hashim2025}, revealing a rich tapestry associated with electron exchange, time-delayed processes and a combination of electron exchange and time delays. However, a unified analytical framework that encompasses more than one excitation channel is lacking.

A key limitation in detangling multichannel interference arises from the huge number of interfering processes. Even for a single channel, there exist numerous phase differences leading to intricate interference patterns. When more than one excitation channel is present, the overwhelming number of phase differences requires additional resources and a complementary strategy, as the analytical tools employed in our previous work no longer suffice to systematically study interference. Furthermore, inter-channel interference depends strongly on the geometry and bound-state energies of the target. There are at least two channels involved, whose interplay may be difficult to map into the resulting patterns. The aim of the present work is to develop a systematic toolkit for identifying, classifying, and quantifying inter-channel interference in RESI whenever it occurs. This toolkit can be transferred to a myriad scenarios within and outside strong-field and attosecond physics, for which intermediate excited states play a role. Examples are HHG
multichannel interference \cite{Smirnova2009,Augstein2012,Hassouneh2014}, RABITT \cite{Paul2001,Dahlstrom2012,Agostini2024} and interferometric spectroscopy, laser-driven dynamics in molecules \cite{Nisoli2017}, and multiphoton ionisation in solids \cite{Ghimire2018,Kruchinin2018}.

Here, we draw upon tools often employed in other fields such as computer vision \cite{Rubner2000, Zhang2020emd}, particle physics \cite{Komiske2019}, biology \cite{Orlova2016} and chemistry \cite{Wang2023emd} to compare the 2D RESI probability distributions rigorously by defining a statistical-distance metric. To our knowledge, within the field of attosecond science, such measures are rarely deployed, with PMDs often compared qualitatively. Typically, when systematic measures are applied, the focus is on assessing the directionality of a single PMD \cite{Landsman2018,Venzke2020}. Widespread examples are the asymmetry parameter, used for instance to characterize left-right asymmetry in streaking experiments, and the photoelectron circular dichroism, widely used in the context of chiral molecules \cite{Bowering2001,Beaulieu2017,Harvey2018,Habibovic2024}. In particular, alternatives to PECD are currently being sought and proposed in order to increase sensitivity. For new classes of enantio-sensitive observables see \cite{Ordonez2023}. The framework described in this present work could aid interpretation of time-resolved photoelectron distributions in molecules, especially when multiple electronic or vibrational channels overlap. It provides new tools for designing control schemes in coherent control or shaped-pulse chemistry. Such tools and observables are sought after in the study of ultrafast chirality, which has applications in the imaging of biomolecules.

We also extend the toolbox of RESI interference conditions by deriving analytic phase differences for multi-channel interference with arbitrary laser fields and classify the distinct interchannel interference types. Using the example of the few-cycle pulse and the simplest case of two-channel interference, we address the question of why some two-channel combinations lead to richer interference patterns than others. What makes two channels comparable? How do differences in excitation energy, relative channel intensities, and the spatial overlap of the excited states influence the resulting two-electron photoelectron momentum distributions? Are some types of interference more sensitive to these factors than others? We use statistical measures, such as the Earth Mover's distance to address these open questions, and show how temporal delays and field-induced phase shifts shape the resulting interference signatures in the correlated electron momentum distributions.
This provides both physical insight and quantitative criteria for when such effects become observable.

Beyond its immediate application to RESI, the methods discussed in the present work address a wider question across many areas of physics: how can one understand the extent to which different quantum pathways contribute to the underlying dynamics? similar questions occur in high-harmonic generation with multiple channels, RABBITT and interferometric spectroscopy, laser-driven dynamics in molecules, multiphoton ionization in solids and more generally in quantum systems where competing pathways coherently contribute to measured distributions. While such problems are ubiquituous, attosecond and strong field physics have so far relied predominantly on qualitative ad hoc comparisons. The EMD provides a more precise tool than, for example, the asymmetry parameter which is widely used in attosecond streaking \cite{Venzke2020}.

More broadly, the methods developed here can inform quantum sensing and simulation by quantifying how interference affects both sensitivity and error accumulation. For quantum sensing, it is important to quantify uncertainties from quantum interference. In \cite{Maxwell2021a}, classical and quantum Fisher information were used to assess sensitivity in strong-field ionization. Because the Fisher information measures how sensitively a probability distribution changes with respect to a parameter, whenever several quantum pathways contribute coherently to an observable (e.g. PMDs, HHG spectra, etc.), their interference is expected to affect that  sensitivity. Earth Mover’s Distance (EMD) or related metrics detect channel-dominance changes, and thus can be employed in conjunction with quantum-sensing tools. 
Similar considerations apply to quantum simulators where interference from competing channels may amplify sensitivity to small parameter changes, but also render the system more fragile and prone to error accumulation. Thus, one must detect when the distribution has changed because the relative channel contributions have shifted. The EMD provides an ideal diagnostic tool in this setting, as it quantifies global redistributions of probability that arise when channel balance or channel phases drift. For recent quantum generalizations of the EMD see \cite{Kiani2022}.

This article is organized as follows. In Sec.~\ref{sec:backgd}, we first review the SFA transition amplitude and saddle-point equations for RESI. Sec.~\ref{sec:2electrondensity} provides a background on how one can sum over channels, events and exchange. In Sec.~\ref{sec:expinterf}, we revisit and adapt the arbitrary-field intrachannel interference conditions for interchannel interference between any two excitation channels. We discuss the expected interference patterns and their location in the $p_{1\parallel}p_{2\parallel}$ parallel momentum plane. Subsequently, in Sec.~\ref{sec:momdists}, we present the PMDs and systematically investigate the different physical factors affecting the shapes and contrast of the interchannel interference. We also analyse the different types of interchannel quantum interferences. Finally, in Sec.~\ref{sec:conclusions}, we state our conclusions and provide further discussion of the implications of the present work on the broader physics community. Unless otherwise stated, atomic units are employed.

\section{Background}
\label{sec:backgd}
\subsection{Transition amplitude}
The SFA  transition amplitude for RESI and an arbitrary excitation channel $\mathcal{C}$ reads
\begin{eqnarray}
&&M^{(\mathcal{C})}(\mathbf{p}_{1},\mathbf{p}_{2})=\hspace*{-0.2cm}\int_{-\infty }^{\infty
}dt\int_{-\infty }^{t}dt^{^{\prime }}\int_{-\infty }^{t^{\prime
}}dt^{^{\prime \prime }}\int d^{3}k  \notag \\
&&\times V^{(\mathcal{C})}_{\mathbf{p}_{2}e}V^{(\mathcal{C})}_{\mathbf{p}_{1}e,\mathbf{k}g}V^{(\mathcal{C})}_{\mathbf{k}%
	g}\exp [iS^{(\mathcal{C})}(\mathbf{p}_{1},\mathbf{p}_{2},\mathbf{k},t,t^{\prime },t^{\prime
	\prime })],  \label{eq:Mp}
\end{eqnarray}
where
\begin{eqnarray}
&&S^{(\mathcal{C})}(\mathbf{p}_{1},\mathbf{p}_{2},\mathbf{k},t,t^{\prime },t^{\prime \prime
})=  \notag \\
&&\quad E^{(\mathcal{C})}_{\mathrm{1g}}t^{\prime \prime }+E^{(\mathcal{C})}_{\mathrm{2g}}t^{\prime
}+E^{(\mathcal{C})}_{\mathrm{2e}}(t-t^{\prime })-\int_{t^{\prime \prime }}^{t^{\prime }}%
\hspace{-0.1cm}\frac{[\mathbf{k}+\mathbf{A}(\tau )]^{2}}{2}d\tau  \notag \\
&&\quad -\int_{t^{\prime }}^{\infty }\hspace{-0.1cm}\frac{[\mathbf{p}_{1}+%
	\mathbf{A}(\tau )]^{2}}{2}d\tau -\int_{t}^{\infty }\hspace{-0.1cm}\frac{[%
	\mathbf{p}_{2}+\mathbf{A}(\tau )]^{2}}{2}d\tau  \label{eq:singlecS}
\end{eqnarray} 
is the semiclassical action. 
Equations~\eqref{eq:Mp} and \eqref{eq:singlecS} have been derived in detail in \cite{Shaaran2010,Shaaran2010a} and 
correspond to a process in which an electron, initially bound in a state of energy $-E^{(\mathcal{C})}_{1g}$, is freed in the continuum at a time $t^{\prime\prime}$. Subsequently, at a time $t^{\prime}$, it returns to its parent ion with intermediate momentum $\mathbf{k}$ and excites a second electron from a bound state with energy  $-E^{(\mathcal{C})}_{2g}$ to a state with energy  $-E^{(\mathcal{C})}_{2e}$. The first electron then leaves, reaching the detector with the final momentum $\mathbf{p}_1$. The second electron is freed at a later time $t$, and has final momentum $\mathbf{p}_2$.  One should note that the RESI action is factorizable, although the transition amplitude is not. Thus, electron-electron correlation is accounted for via the time ordering: recollision of the first electron must happen before ionization of the second electron \cite{Shaaran2012}.

In the SFA, all information about the target geometry and the interactions is embedded in the 
prefactors  $V^{(\mathcal{C})}_{\mathbf{k}g}$, $V^{(\mathcal{C})}_{\mathbf{p}_1e,\mathbf{k}g}$ and $V^{(\mathcal{C})}_{\mathbf{p}_2e}$ \cite{Shaaran2010,Rook2024}. These are associated with the ionization of the first electron, the recollision-excitation process and the tunnel ionization of the second electron, respectively.  The expressions are general and the superscript $(\mathcal{C})$ make them easily adaptable to coherent superpositions of channels and bound states.

The first electron ionization prefactor reads
    \begin{equation}\label{eq:pre3}
V^{(\mathcal{C})}_{\mathbf{k}g} = \bra{\mathbf{k}}V\ket{\psi_{1g}^{(\mathcal{C})}},
\end{equation}
where $V(\mathbf{r}_1)$ is the neutral atom's binding potential, and $\braket{\mathbf{r}_1|\psi_{1g}^{(\mathcal{C})}}=\psi_{1g}^{(\mathcal{C})}(\mathbf{r}_1)$ is the ground-state wave function for the first electron.

The excitation prefactor for the second electron is given by 
\begin{equation}
\label{eq:Vp1ekg}V^{(\mathcal{C})}_{\mathbf{p}_1e,\mathbf{k}g}\hspace*{-0.1cm} = \hspace*{-0.1cm} \bra{\mathbf{p}_1,\psi_{2e}^{(\mathcal{C})}}V_{12}
\ket{\mathbf{k},\psi_{2g}^{(\mathcal{C})}}, 
\end{equation}
where
\begin{equation}
    V_{12}(\mathbf{p}_1 - \mathbf{k}) =  \frac{1}{(2\pi)^{3/2}}\int d^3rV_{12}(\mathbf{r})\exp[-i(\mathbf{p}_1-\mathbf{k})\cdot\mathbf{r}]
\end{equation}
is the electron-electron interaction in momentum space, $\mathbf{r} = \mathbf{r}_1-\mathbf{r}_2$, and $V_{12}(\mathbf{r})$, taken to be of contact type, describes the interaction by which the second electron is excited.  The wave functions $\braket{\mathbf{r}_2|\psi_{2e}^{(\mathcal{C})}}=\psi_{2e}^{(\mathcal{C})}(\mathbf{r}_2)$ and $\braket{\mathbf{r}_2|\psi_{2g}^{(\mathcal{C})}}=\psi_{2g}^{(\mathcal{C})}(\mathbf{r}_2)$ are associated with the excited and ground states of the second electron, respectively.
Finally, the second electron ionization prefactor reads
\begin{equation}
    V^{(\mathcal{C})}_{\mathbf{p}_2e} = \bra{\mathbf{p}}V_{\mathrm{ion}}\ket{\psi_{2e}^{(\mathcal{C})}}  
    \label{eq:Vp2e}
\end{equation}
where $V_{\mathrm{ion}}(\mathbf{r}_2)$ is the potential of the singly ionized target, describes the ionization of the second electron. 
In our calculations, we employ hydrogenic wave functions $\psi_{nlm}(\textbf{r}) = R_{nl}(r)Y^m_l(\theta, \phi)$ for the electronic bound states. The explicit expressions for these prefactors can be found in the appendix of \cite{Hashim2024}. We have taken the magnetic quantum number $m=0$ to faciliate comparison with previous works \cite{Maxwell2016} and with existing results in literature \cite{Hao2014}. 

We consider the ionization prefactors in the velocity gauge to avoid the bound-state singularities arising due to the saddle-point equation \eqref{eq:sp4} describing the ionization of the second electron, which leads to vanishing denominators in $V_{\textbf{p}_2e}$ for a hydrogenic basis. Means to overcome this singularity, as well as a detailed discussion, are given in \cite{Shaaran2010}. 
The additional momentum shifts resulting from the length-gauge prefactors will not play an important role for the ionization prefactor \eqref{eq:Vp2e}, due to the vector potential at the prevalent tunnel ionization times being vanishingly small for the field in question. Throughout, these shifts
will cancel out for the excitation prefactor \eqref{eq:Vp1ekg} for the field in question - see \cite{Shaaran2010} for a detailed discussion. For other field shapes, such as bichromatic fields with driving waves of comparable strengths,  due to a non-vanishing vector potential at the ionization times, this no longer holds for $V_{\mathbf{p}_2e}$ and 
it is necessary to consider length-gauge prefactors \cite{Hashim2024b, Hashim2025}.

\subsection{Saddle-point equations}
\label{sec:spe}
The integrals in the transition amplitude Eq.~\eqref{eq:Mp} are calculated using the saddle-point method, which seeks variables $\textbf{k}, t'', t'$ and $t$ such that the action is stationary. This leads to the saddle-point equations
\begin{equation}\label{eq:sp1}
[\mathbf{k}+\mathbf{A}(t'')]^2 = -2E^{(\mathcal{C})}_{1g},
\end{equation}
\begin{equation}\label{eq:sp2}
\mathbf{k} = -\frac{1}{t'-t''}\int_{t''}^{t'}d\tau \mathbf{A}(\tau)
\end{equation}
\begin{equation}\label{eq:sp3}
[p_{1\parallel}+A(t')]^2+[\mathbf{p}_{1\perp}]^2 = [\mathbf{k}+\mathbf{A}(t')]^2 - 2(E^{(\mathcal{C})}_{2g}-E^{(\mathcal{C})}_{2e})
\end{equation}
and 
\begin{equation}\label{eq:sp4}
[p_{2\parallel} + A(t)]^2+[\mathbf{p}_{2\perp}]^2 = -2E^{(\mathcal{C})}_{2e}.
\end{equation}
Eqs.~\eqref{eq:sp1} and \eqref{eq:sp4} give the conservation of energy upon tunnel ionization for the first and second electrons, respectively. Eq.~\eqref{eq:sp2} constrains the intermediate momentum so that the first electron returns to the site of its release, and Eq.~\eqref{eq:sp3} gives the rescattering event in which the first electron gives part of the kinetic energy $[\mathbf{k}+\mathbf{A}(t')]^2/2$ to excite the second electron.  
We denote $p_{n\parallel}$ and $\mathbf{p}_{n\perp}$, $(n=1,2)$, the momentum components parallel and perpendicular to the laser-field polarization, respectively, where $\mathbf{p}_{n\perp}$ is a two-dimensional vector spanning the perpendicular-momentum plane. The integrals are then approximated by sums over these stationary variables. This allows linking the saddle-point solutions to the times associated with electron trajectories: their real parts, $\mathrm{Re}[t], \mathrm{Re}[t'], \mathrm{Re}[t'']$, are directly related to the classical recollision and ionization times, while $\mathrm{Im}[t], \mathrm{Im}[t'']$ are loosely associated with the instantaneous tunneling probability $\sim \exp [-2\mathrm{Im}[S]]$ of the second and first electron, respectively \cite{Shaaran2012a}. The real parts of the ionization and rescattering times are close to the field maxima and zero crossings, respectively \cite{Shaaran2010,Hashim2024}. 

For the second electron, the saddles are well separated in all momentum regions and thus the standard saddle point approximation can be applied, while, for the first electron, pairs of saddles must be considered collectively using the uniform approximation in \cite{Faria2002}. 
Eq.~\eqref{eq:sp3} relates the kinetic energy $E_k (t',t'')$ of the returning electron to the energy difference $\Delta E^{(\mathcal{C})} = E_{2g}^{(\mathcal{C})} - E_{2e}^{(\mathcal{C})}$ between the ground and excited states of the second electron. 
The saddle-point equation defines a sphere in the momentum space of the first electron, $(p_{1\parallel}, \mathbf{p}_{1\perp})$, with radius $ \sqrt{2[E_k(t',t'') - \Delta E^{(\mathcal{C})}]} $. A real radius exists when $ E_k(t',t'') > \Delta E^{(\mathcal{C})} $ suggesting a possible classical counterpart for Eq.~\eqref{eq:sp3}.  By setting $p_{1\perp}=\mathbf{0}$, an upper bound on $p_{1\parallel}$ is established so that this condition is fulfilled, defining the \textit{classically allowed region} (CAR) for rescattering. Beyond this region, the probability density associated with the first electron decays exponentially. This concept has been introduced in our previous publications \cite{Shaaran2010, Rook2024}. 
Furthermore, the saddle-point equations state that RESI may be viewed as two time-ordered processes - rescattered inelastic above-threshold ionization (ATI) for the first electron and direct ATI for the second electron.

\subsection{Two-electron probability density}
\label{sec:2electrondensity}
The quantity of interest is the RESI two-electron probability density as a function of the momentum components $p_{n\parallel} (n=1,2)$ parallel to the driving-field polarization, given by
\begin{align}
	\mathcal{P}(p_{1\parallel},p_{2\parallel})= \int\int d^2 p_{1\perp}d^2 p_{2\perp}\mathcal{P}(\mathbf{p}_{1},\mathbf{p}_{2}), \label{Eq:Channels}
\end{align}
where $\mathcal{P}(\mathbf{p}_{1},\mathbf{p}_{2})$ is the fully resolved two-electron momentum probability density, and the transverse momentum components have been integrated over. Several issues must be taken into consideration upon calculation of this probability density. The first is electron exchange. Due to the indistinguishability of the electrons, Eq.~\eqref{eq:Mp} must be symmetrized upon $\mathbf{p}_1\leftrightarrow\mathbf{p}_2$. Secondly, there will be several `events' within the pulse. Thirdly, in a real target, there will be several excitation channels, each of which will have an associated transition amplitude. Thus, within the saddle-point approximation, the overall RESI amplitude must contain sums over (i) symmetrization related to electron indistinguishability which will occur for each pair of excitation and ionization times, (ii) the events within a pulse and (iii) transitions involving different excitation channels. Quantum mechanically, all these contributions add up coherently.  
The fully coherent sum over events, channels and symmetrization reads
\begin{equation}
\mathcal{P}_{(\mathrm{ccc})}(\mathbf{p}_{1},\mathbf{p}_{2})=\left|\sum_{\varepsilon}\sum_{\mathcal{C}}M_{\varepsilon}^{(\mathcal{C})}(\mathbf{p}_1,\mathbf{p}_2)+M_{\varepsilon}^{(\mathcal{C})}(\mathbf{p}_2,\mathbf{p}_1)\right|^2, 
\label{eq:fullcoherent}
\end{equation}
where the symbols $\varepsilon$ and $\mathcal{C}$ denote event and channel respectively,  the symmetrization is indicated by $\mathcal{S}$, and the subscript $(\mathcal{S}\mathcal{E}\mathcal{C})=(\mathrm{ccc})=$(coherent-coherent-coherent) indicates full coherence. Throughout, we use the notation 
$\mathcal{P}^{(\mathcal{C}_1, \mathcal{C}_2,..., \mathcal{C}_n)}_{(s \epsilon \mathcal{C})}$ the indices are associated with the  symmetrization, event and channel respectively. Thus, $\mathcal{P}^{(\mathcal{C}_1, \mathcal{C}_2, ...\mathcal{C}_n)}_{(c c c)}$ states that the sum considered in the probability density is coherent over the pulse events, symmetrization and channels, and involves the $\mathcal{C}_1, \mathcal{C}_2,...\mathcal{C}_n$-th channels. 

However, since we are interested in detangling quantum interference, we will construct the two-electron probability density in several ways depending on the question we wish to address.
A summary of the possible different sums for the events, channel and symmetrization is provided in Table \ref{tab:possiblesums}. For simplicity,  we focus on two-channel interference, although the approach may be generalized to an arbitrary number of excitation channels. 

\begin{table}[ht]
\centering
\begin{adjustbox}{max width=280pt}
\begin{tabular}{cc}
\hline
\textbf{Process} $\mathcal{SEC}$ &  \textbf{$\mathcal{P_{SEC}}(p_{1}, p_{2})$} \\
\hline \hline
(a) ccc & $|\sum_{\varepsilon}\sum_{\mathcal{C}}M_{\varepsilon}^{(\mathcal{C})}(\mathbf{p}_1,\mathbf{p}_2)+M_{\varepsilon}^{(\mathcal{C})}(\mathbf{p}_2,\mathbf{p}_1)|^2$ \\
(b) cic & $\sum_{\varepsilon}\left|\sum_{\mathcal{C}}M_{\varepsilon}^{(\mathcal{C})}(\mathbf{p}_1,\mathbf{p}_2)+M_{\varepsilon}^{(\mathcal{C})}(\mathbf{p}_2,\mathbf{p}_1)\right|^2$ \\
(c) icc & $\left|\sum_{\varepsilon}\sum_{\mathcal{C}}M_{\varepsilon}^{(\mathcal{C})}(\mathbf{p}_1,\mathbf{p}_2)\right|^2+\left|\sum_{\varepsilon}\sum_{\mathcal{C}}M_{\varepsilon}^{(\mathcal{C})}(\mathbf{p}_2,\mathbf{p}_1)\right|^2$\\
(d) iic & $\sum_{\mathcal{\varepsilon}}|\sum_{\mathcal{C}}M_{\varepsilon}^{(\mathcal{C})}(\mathbf{p}_1,\mathbf{p}_2)|^2+|\sum_{\mathcal{C}}M_{\varepsilon}^{(\mathcal{C})}(\mathbf{p}_2,\mathbf{p}_1)|^2$\\
(e) cci & $\sum_{\mathcal{C}}\left|\sum_{\varepsilon}M_{\varepsilon}^{(\mathcal{C})}(\mathbf{p}_1,\mathbf{p}_2)+M_{\varepsilon}^{(\mathcal{C})}(\mathbf{p}_2,\mathbf{p}_1)\right|^2$ \\
(f) cii & $\sum_{\varepsilon}\sum_{\mathcal{C}}|M_{\varepsilon}^{(\mathcal{C})}(\mathbf{p}_1,\mathbf{p}_2)+M_{\varepsilon}^{(\mathcal{C})}(\mathbf{p}_2,\mathbf{p}_1)|^2$\\
(g) ici & $\sum_{\mathcal{C}}|\sum_{\varepsilon}M_{\varepsilon}^{(\mathcal{C})}(\mathbf{p}_1,\mathbf{p}_2)|^2+|\sum_{\varepsilon}M_{\varepsilon}^{(\mathcal{C})}(\mathbf{p}_2,\mathbf{p}_1)|^2$\\
(h) iii & $\sum_{\varepsilon}\sum_{\mathcal{C}}\left|M_{\varepsilon}^{(\mathcal{C})}(\mathbf{p}_1,\mathbf{p}_2)\right|^2+\left|M_{\varepsilon}^{(\mathcal{C})}(\mathbf{p}_2,\mathbf{p}_1)\right|^2$ \vspace*{0.1cm} \\ \hline \hline
\end{tabular}
\end{adjustbox}
\caption{ Possible ways of constructing probability densities for a pulse, in terms of events, channels and symmetrization. The letters c and i indicate coherent and incoherent sums, respectively regarding symmetrization (denoted by $\mathcal{S}$), event ($\mathcal{E}$) and channel ($\mathcal{C}$). \label{tab:possiblesums} }
\end{table}

Furthermore, in this work, we also consider the interference of a specific event with the symmetrized counterpart of a time-delayed event. In this case, assuming two events $\varepsilon$ and $\varepsilon '$, $\varepsilon \neq \varepsilon'$, and summing these events pairwise, the corresponding coherent sum reads
\begin{equation}
    \mathcal{P}_{(\mathrm{ccc, \varepsilon\varepsilon'})}(\mathbf{p}_{1},\mathbf{p}_{2})=\sum_{\varepsilon, \varepsilon', \mathcal{C}}\left|M_{\varepsilon}^{(\mathcal{C})}(\mathbf{p}_1,\mathbf{p}_2)+M_{\varepsilon'}^{(\mathcal{C})}(\mathbf{p}_2,\mathbf{p}_1)\right|^2.
    \label{eq:combinedP}
    \end{equation}
    Eq.~\eqref{eq:combinedP} differs from process (a) in Table \ref{tab:possiblesums}, which is denoted by ccc but considers the same events.

Here, we focus on interchannel interference so only probability densities with coherent sums over channels will be considered [processes (a)-(d) in Table.~\ref{tab:possiblesums}]. Process (a) is the fully coherent sum over all events and channels. In process (b), the symmetrization is done coherently while the events are summed over incoherently and vice-versa in process (c). These processes will detangle the additional interference arising from exchange and temporal shifts, respectively and will be referred to as `exchange-channel' and `event-channel' interference therein.  Process (d) considers the case where both symmetrization and events are summed over incoherently, and only the channels are summed coherently. We will therefore refer to this process as `channel-only' interference. 

Often, it is necessary to carry out partial sums in which specific events are considered pairwise or in which we look at an individual symmetrized event. We indicate differences of probability densities computed by different means by $\mathcal{P}_\text{diff}(p_{1\parallel}, p_{2\parallel})$ for simplicity in the figure axes, but are more specific about which differences we consider in the captions and discussions. 

\subsection{Target considerations}
\label{sec:target} 
Here, we will consider Argon, for which there are six main excitation channels. The absolute values of the ionization potential associated with the neutral and singly ionized target are $E^{(\mathcal{C})}_{1g}=0.58$ a.u and $E^{(\mathcal{C})}_{2g}= 1.016$ a.u., respectively for all channels $\mathcal{C}=1$ to $6$. The electronic configuration before excitation is $1s^22s^22p^63s^23p^5$ for all channels, but in channel 1 a hole is created in $3s$ upon recollision $(3s3p^6)$. For $\mathcal{C}=2$ to 6, the second electron electron is excited from the outer shell.  These channels are given in Table \ref{tab:channels}. The fine structure has been neglected as typical fine-structure energy splittings for low-lying levels of Ar and Ar$^+$ are of the order of $\Delta E_{\rm fs}\sim 10^{-3} - 10^{-2}~{\rm a.u.}$, which is tiny in comparison with the remaining energy scales in this work. In order for all these channels to interfere, it is necessary that, in addition to final momentum states, the final ionic state be the same, i.e., $3s^23p^4$. However, we have frozen the ion in our computations, thus neglecting the recombination process from the intermediate state  $3s3p^5$ in channel 1. Because the present work serves more as proof of concept of the statistical methods used to quantify interference and the hierarchy of channel contributions, it is legitimate to neglect recombination process. However, it should be incorporated in a more realistic comparison.  

\begin{table}[] 
\begin{tabular}{cccc}
\hline \hline
 Channel & Excited-state configuration  &  $E^{(\mathcal{C})}_{2e}$ (a.u.)& Maxima   \vspace*{0.1cm} \\ \hline
 1& $3s3p^6 (3s \rightarrow 3p)$ &  0.52 & $d$   \\
 2& $3p^53d (3p \rightarrow 3d)$ &  0.41 & $a,d$  \\
 3& $3p^54d (3p \rightarrow 4d)$ &  0.18 & $ a,d$  \\
 4&  $3p^54s (3p \rightarrow 4s)$  &  0.40&  $a$ \\
 5& $3p^54p (3p \rightarrow 4p) $& 0.31 &  $d$ \\
 6& $3p^55s (3p \rightarrow 5s)$ & 0.19 &   $a$ \vspace*{0.1cm} \\ \hline \hline
\end{tabular}
\caption{Relevant excitation channels for $\mathrm{Ar}^+$, in order of increasing principal and orbital quantum numbers. From left to right, the first column gives the number associated with the channel, the second column states the electronic configuration and the excitation pathway, the third column provides the excited-state energy in atomic units, and the fourth column specifies whether the resulting distributions have maxima at the axes, diagonals, or both. The letters $a$ and $d$ in the fourth column stand for axes and diagonals, respectively. \label{tab:channels}}
\end{table}

Each excitation channel contributes to the PMDs via its bound-state energy and bound-state geometry. The bound-state energies affect the saddle-point equations. Eqs.~(\ref{eq:sp1}) and (\ref{eq:sp4}) are associated with the ionization probability of the first and second electron. The more tightly bound the electrons are, the lower their tunneling probability will be. Eq.~\eqref{eq:sp3}, associated with rescattering, determines the classically allowed region and is affected by the energy gap $\Delta E^{(\mathcal{C})}= (E^{(\mathcal{C})}_{2g}-E^{(\mathcal{C})}_{2e})$. Loosely bound excited states imply large excitation gaps and therefore a small, or inexistent, classically allowed region, causing a suppression in the first electron's probability density. However, loosely bound states lead to a high tunneling probability for the second electron. These competing effects determine the relative contributions of individual channels to the overall PMDs. The prefactors introduce biases associated with the bound-state geometry. The shapes of $\mathcal{P}(p_{1\parallel},p_{2\parallel})$ are mainly determined by $V_{\mathbf{p}_2e}$ \cite{Shaaran2010,Maxwell2015}, while the remaining prefactors exert a subtler influence. This is due to $\mathbf{p}_2$ being time-independent and the vector potential $\mathbf{A}(t)$ being vanishingly small at the dominant ionization times determined by Eq.~\eqref{eq:sp4}, while the time dependence of the intermediate momentum $\mathbf{k}$ [see Eq.~\eqref{eq:sp2}] blurs the nodes. The fourth column of Table \ref{tab:channels} roughly indicates the maxima associated with $s$ (axes), $p$ (diagonals) and $d$ (axes and diagonals) excited states, although their actual mapping is more complicated \cite{Hashim2024}.

For coherent sums over \textit{multiple} excitation channels, one must consider the energy difference $E_\text{diff}^{(\mathcal{C}_n, \mathcal{C}_m)}= \Delta E^{(\mathcal{C}_n)} - \Delta E^{(\mathcal{C}_m)}$. Although the excitation gaps $\Delta E^{(\mathcal{C}_n)}$ affect the width and intensity of the $n-$th single-channel PMD, $E_\text{diff}^{(\mathcal{C}_n,\mathcal{C}_m)}$ determines the relative strength of the contributions of any two channels in the full multichannel PMD, which is expected to contain interference features arising from the contributing channels. For instance, transitions such as $3s\rightarrow3p$ and $3p\rightarrow4p$ involving $p$ states are expected to contribute primarily along the diagonals, with nodes along the axes. However, even when the orbital angular momentum quantum number $l$ of the excited state remains fixed, variations in the principal quantum number $n$ can alter the shape of the excitation prefactor $V_{\mathbf{p}_2e}$ (and thus its mapping). Consequently, the node and maxima in the two-channel PMDs may deviate from the typical $p$-state distribution.

When the excited states differ in orbital angular momentum $l$ one may observe maxima along both the axes and the diagonals—for instance, in multichannel PMDs with contributions from the $3p\rightarrow4s$ and $3p\rightarrow4p$ transitions. Predicting the precise shape of the resulting two-channel PMDs is challenging, as the relative intensity of the individual single-channel PMDs strongly influences which features dominate the total distribution. In summary, the final shape and intensity of the multichannel PMD are determined by a combination of factors: the difference in excitation gap $E_\text{diff}$  (whose effect is nontrivial due to competing contributions from the first electron’s residual kinetic energy and the second electron’s tunneling probability), the relative intensity of the channels and the characteristic shapes of the contributing single-channel PMDs.

\subsection{Momentum constraints}
\label{sec:field}
Next, we recall the regions in the $p_{1\parallel} p_{2\parallel}$ plane occupied by the RESI distributions from the few-cycle pulse employed in this work, which are detailed in  \cite{Hashim2024}.  Sketching these regions uses: (a) the real parts of the saddle-point solutions, which can be associated with classical times; (b) ionization being most probable around a field extremum and rescattering around a field zero crossing, so the distributions will be centered around these times. These estimates dictate that the final momentum of both electrons will be located around  $(p_{ 1\parallel},p_{ 2\parallel})=( -A(t'),-A(t))$. The momentum regions for which the PMDs are significant are approximately determined by the direct and rescattered ATI cutoff energies for the second and first electron, respectively \cite{Shaaran2010,Shaaran2010a}. Because the rescattered ATI cutoff energy is much higher than that of direct ATI\footnote{For a monochromatic field or long enough pulses, the direct (rescattered) ATI cutoff energy is $2U_p$ (10$U_p$), where $U_p$ is the ponderomotive energy. These numbers relate to the maximal kinetic energy a direct or rescattered ATI electron may acquire from the external field, are approximate for few cycle pulses. Bi- or polychromatic fields will lead to different cutoff laws, but the final kinetic energy of the first electron will always be higher. For NSDI RESI, there is an inelastic collision, so that the energy gap $E_{2g}-E_{2e}$ should be subtracted from the rescattered ATI cutoff.  }, the width and the length of the distributions are determined by the second and the first electron, respectively. The extension of the momentum region occupied will depend on the electron's kinetic energy upon return at each rescattering event. This region may be large, small, or even have no classical counterpart. Electron indistinguishability dictates that there will also be events whose amplitudes are centered at $(p_{ 1\parallel},p_{2\parallel})=( -A(t),-A(t'))$. The shape and symmetry of the RESI distributions will depend on that of the driving field, with the field-determined symmetry being retained for fully incoherent probability densities. This issue has been addressed in \cite{Rook2022} for ATI and in \cite{Hashim2024b} for RESI in bichromatic fields using group-theoretical arguments. Thus, we will only outline the key points below.

We employ a linearly polarized few-cycle pulse $\mathbf{E}(t)=-d\mathbf{A}(t)/dt$, whose vector potential is determined by
\begin{equation}
\mathbf{A}(t)=2 A_0\sin^2\left(\frac{\omega t}{2 N}\right)\sin{(\omega t + \phi)}\hat{e}_z,
    \label{eq:Apulse}
\end{equation}
where $A_0$ is the vector-potential amplitude, $N$ is the number of cycles, $\omega$ is the field frequency, $\phi$ the carrier-envelope phase.  Throughout, we have made the approximation  $A_0 = 2\sqrt{U_p}$, where $U_p$ is the ponderomotive energy. This is exact for a monochromatic linearly polarized field, but not for a few-cycle pulse.  It was determined that the most dominant events in the pulse are those in the center. We only consider the ionization event for the second electron immediately following the rescattering of the first electron, as the probability of later events is smaller due to bound-state depletion and therefore less important. 

To determine the momentum regions occupied by the multichannel PMD, we must first consider the constraints for a single channel. Similar to the procedure in \cite{Maxwell2015} for a monochromatic wave, neighboring events within the pulse, displaced by approximately half a cycle, and those present due to electron exchange symmetry will lead to the transition amplitudes $M_l$, $M_u$, $M_r$, and $M_d$ with $M_l(\mathbf{p}_1,\mathbf{p}_2)=M_d(\mathbf{p}_2,\mathbf{p}_1)$ and $M_r(\mathbf{p}_1,\mathbf{p}_2)=M_u(\mathbf{p}_2,\mathbf{p}_1)$ where $M_l$, $M_u$, $M_r$, and $M_d$ refer to a probability amplitude in the left, up, right and down directions, respectively. The subscripts were chosen to highlight the main momentum regions occupied by each amplitude in the $p_{1\parallel}p_{2\parallel}$ plane, whose lengths and widths are defined by the cutoff energies associated with the first and second electron, respectively. Details on these constraints are given in \cite{Shaaran2010a} for a monochromatic driving field and in \cite{Faria2012} for few-cycle pulses. 

\begin{figure*}
    \centering
    \includegraphics[width=\textwidth]{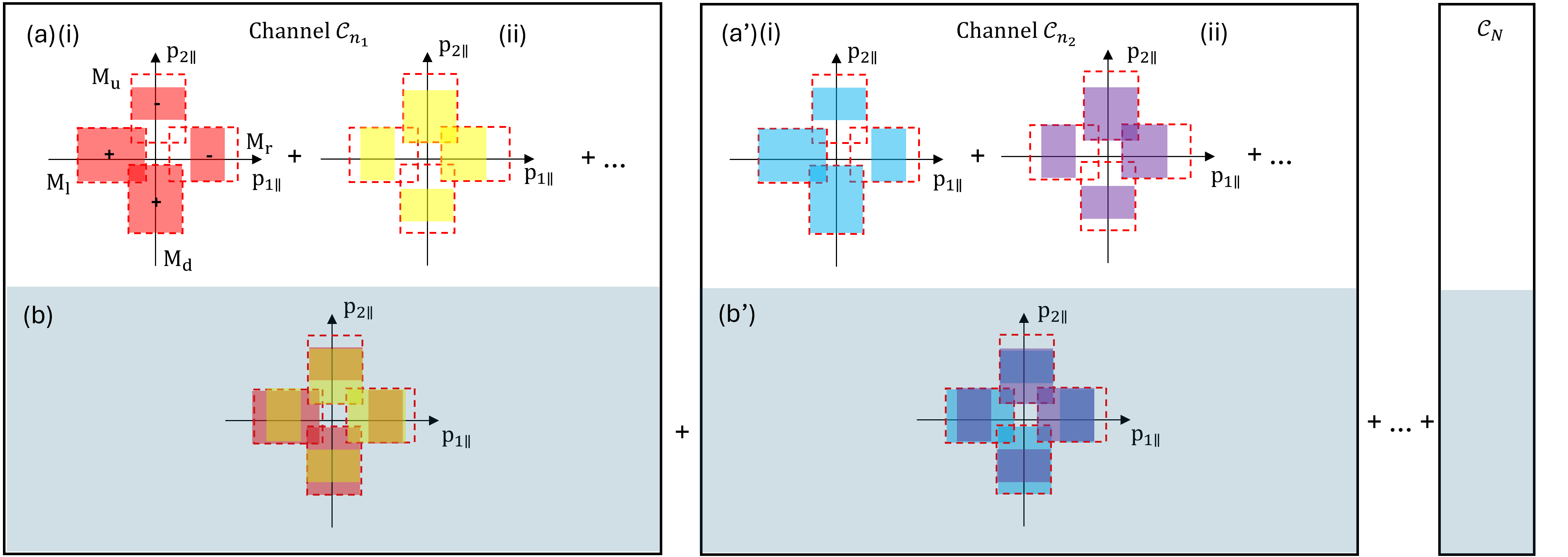}
    \caption{Schematic representation of the momentum-space regions occupied by the RESI transition amplitudes $M_l, M_u, M_r$ and $M_d$ associated with different events within the pulse [panels (a, a')], and those resulting from their coherent superposition [panel (b)] for multiple channels. The negative (positive) signs in panel (ai) indicate the most probable momenta. The dashed rectangles indicate the momentum constraints that would hold for a monochromatic driving field, while the shaded areas show their counterparts for few-cycle pulses. We have used the same color for $M_l$ and $M_d$, and $M_u$ and $M_r$ to highlight the property $M_l(\mathbf{p_1}, \mathbf{p_2)}= M_d(\mathbf{p_2}, \mathbf{p_1})$ and $M_r(\mathbf{p_1}, \mathbf{p_2)}= M_u(\mathbf{p_2}, \mathbf{p_1})$. We assumed that each subpanel in (a) gives the momentum regions occupied by events within a single cycle, and the events in (a)(i) and (a)(ii) are summed coherently.  
    Overlapping shaded regions indicate that quantum interference may occur. We consider two events separated by a half cycle in each plot in panel (a), which eventually interfere, but one can extend to any number of interfering events. The full coherent photoelectron momentum distribution (i.e. summed over all excitation channels, events and symmetrization) will occupy the full momentum space, and will be the sum of the single-channel PMDs shown in panels (a) and (b). This idea can also be extended for multiple channels.}
    \label{fig:schematic1}
\end{figure*}

Figures~\ref{fig:schematic1}(ai)(aii) and (ai')(aii') present schematic representations of the momentum regions associated with the matrix elements for two distinct events within the pulse, corresponding to arbitrary channels $\mathcal{C}_n$ and $\mathcal{C}_m$, respectively.
In the figure, the shaded regions indicate the momentum constraints occupied by each of these transition amplitudes, associated with the ATI cutoff energies for $\mathbf{p}_{n\perp}=\mathbf{0}$. Outside these constraints, the probability density is strongly suppressed as the corresponding transition amplitude has no classical counterpart.

The shaded regions are reflection-symmetric about to the diagonal $p_{1\parallel}=p_{2\parallel}$, which is guaranteed by electron indistinguishability. However, other symmetries are missing. This happens because the pulse given by Eq.~\eqref{eq:Apulse} is not expected to exhibit any of the symmetries of a monochromatic wave, which are: (i) the half-cycle symmetry  $\textbf{A}(t) \neq \pm \textbf{A} (t \pm T/2)$ ; (ii) reflection symmetry about the field extrema; (iii) reflection symmetry about the field zero crossings, together with a reflection about the temporal axis. These three symmetries guarantee fourfold symmetric RESI distributions ($M_l(\mathbf{p}_1,\mathbf{p}_2)=M_r(-\mathbf{p}_1,-\mathbf{p}_2)$ and $M_u(\mathbf{p}_1,\mathbf{p}_2)=M_d(-\mathbf{p}_1,-\mathbf{p}_2)$), which is not the case here (see dashed red lines in the figure). Half-cycle symmetric fields (without necessarily (ii) and (iii) holding) yield RESI distributions that are reflection-symmetric about the anti-diagonal $p_{1\parallel}=-p_{2\parallel}$, which is also broken \cite{Hashim2024b}. 

In addition to that, Fig.~\ref{fig:schematic1} indicates that for each channel, interference will be substantial for the coherent sums $M_{ld}=M_l+ M_d$, $M_{ru}=M_r+ M_u$, $M_{ul}=M_u+ M_l$, and $M_{rd}=M_r+ M_d$.  To calculate intrachannel interference, one can calculate the phase differences associated with them within a single cycle. The sums $M_{ld}$ and $M_{ru}$ will depend on symmetrization only, while for $M_{ul}$ and $M_{rd}$ one must consider a half-cycle shift and a symmetrization. The sums $M_{rl}=M_l+ M_r$ and $M_{ud}=M_u+ M_d$ of events separated by half a cycle will also overlap and interfere, but this interference will be negligible as the probability density is vanishingly small outside the shaded regions (for analytic proofs and explicit computations, using monochromatic fields, see \cite{Shaaran2010} and \cite{Maxwell2015}, respectively). 

This work focuses on interference between events occurring in different channels. Thereby, key questions are what types of interchannel interference are there and under what conditions they are substantial. We will see interference between events depicted in Fig.~\ref{fig:schematic1}(ai), (aii) from channel $C_{n_1}$ with those in Fig.~\ref{fig:schematic1}(ai'), (aii') from channel $C_{n_2}$. The fully coherent sum will involve intrachannel interference as depicted in Fig.~\ref{fig:schematic1}(b),(b'), as well as interference from summing the channels coherently. Therefore, even for interference between just two channels, one expects coherent effects to span the entire momentum region occupied by the contributing single-channel PMDs. If the events occupy very different momentum regions in different channels, for instance due to very different geometries of the excited states, one will see PMDs which span a wider momentum range, but interference effects only where the events overlap according to the present mapping. Channel-only interference between two channels $\mathcal{C}_n$ and $\mathcal{C}_m$ will be substantial for $M_{\mu,\mu}^{(\mathcal{C}_n, \mathcal{C}_m)} = M_\mu^{\mathcal{C}_n} + M_\mu^{\mathcal{C}_m}$ where $\mu=l,r,d,u$ i.e. where the same events in both channels overlap. Channel-exchange will remain substantial for the coherent sums $M_{l,d}^{(\mathcal{C}_n, \mathcal{C}_m)}$ and $M_{r,u}^{(\mathcal{C}_n, \mathcal{C}_m)}$ where now the transition amplitudes belong to different excitation channels. Similarly, channel-temporal will be substantial for the coherent sums $M_{l,r}^{(\mathcal{C}_n, \mathcal{C}_m)}$ and $M_{u,d}^{(\mathcal{C}_n, \mathcal{C}_m)}$. Finally, the channels in question should also be energetically close, and produce PMDs which are comparable in intensity, lest one dominate over the other. 

\subsection{Field-specific effects}

\begin{figure}
    \centering
\includegraphics[width=\columnwidth]{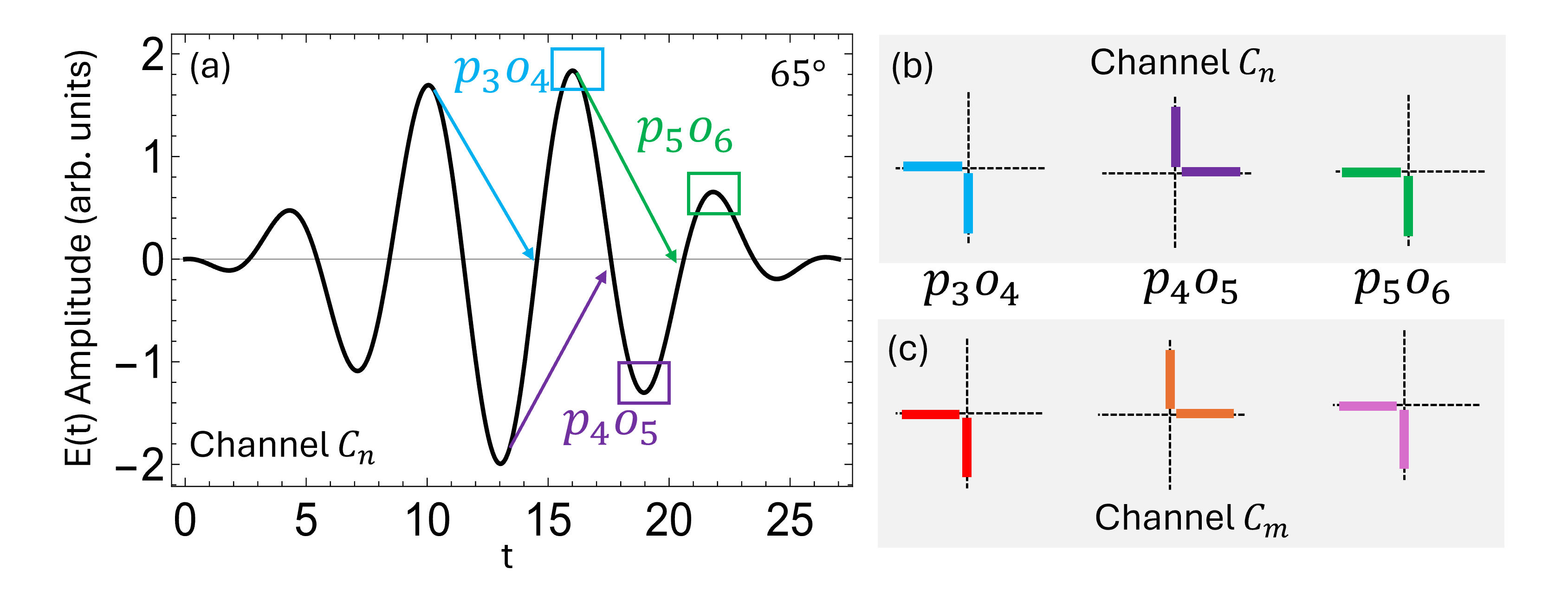}
    \caption{Few-cycle pulse associated with the vector potential (\ref{eq:Apulse}), with peak intensity $I=1.5 \times 10^{14}\mathrm{W/cm}^2$, wavelength $\lambda=800$ nm ($\omega=0.057$ a.u.), $N=4.3$ and carrier-envelope phases $\phi_1=65^{\circ}$ [panel (a)]. The three main events $p_io_j$ towards the center of the pulse are labeled with their corresponding pair and orbit numbers. The classical ionization and return times associated with the pairs of orbits $p_3$, $p_4$, $p_5$ of the first electron are indicated by arrows, and the most relevant ionization times for the orbits $o_j$ of the second electron, with $(j=4,5,6)$, are signposted by rectangles. The initial numbers chosen for the indices $i,j$ refer to the extremum of the field for which the counting starts.  Matching styles and colors have been used for different events $\varepsilon_k=p_io_j$ and the momentum mapping in panel (b), associated with the channel $\mathcal{C}_n$. Panel (b) displaces an analogous mapping for a different channel $\mathcal{C}_m$. To facilitate the interference studies, in this latter panel we have employed different colors, although the sketched PMDs are associated with the same events. }
    \label{fig:pulseshape}
\end{figure}

\begin{figure}
    \centering
    \includegraphics[width=\columnwidth]{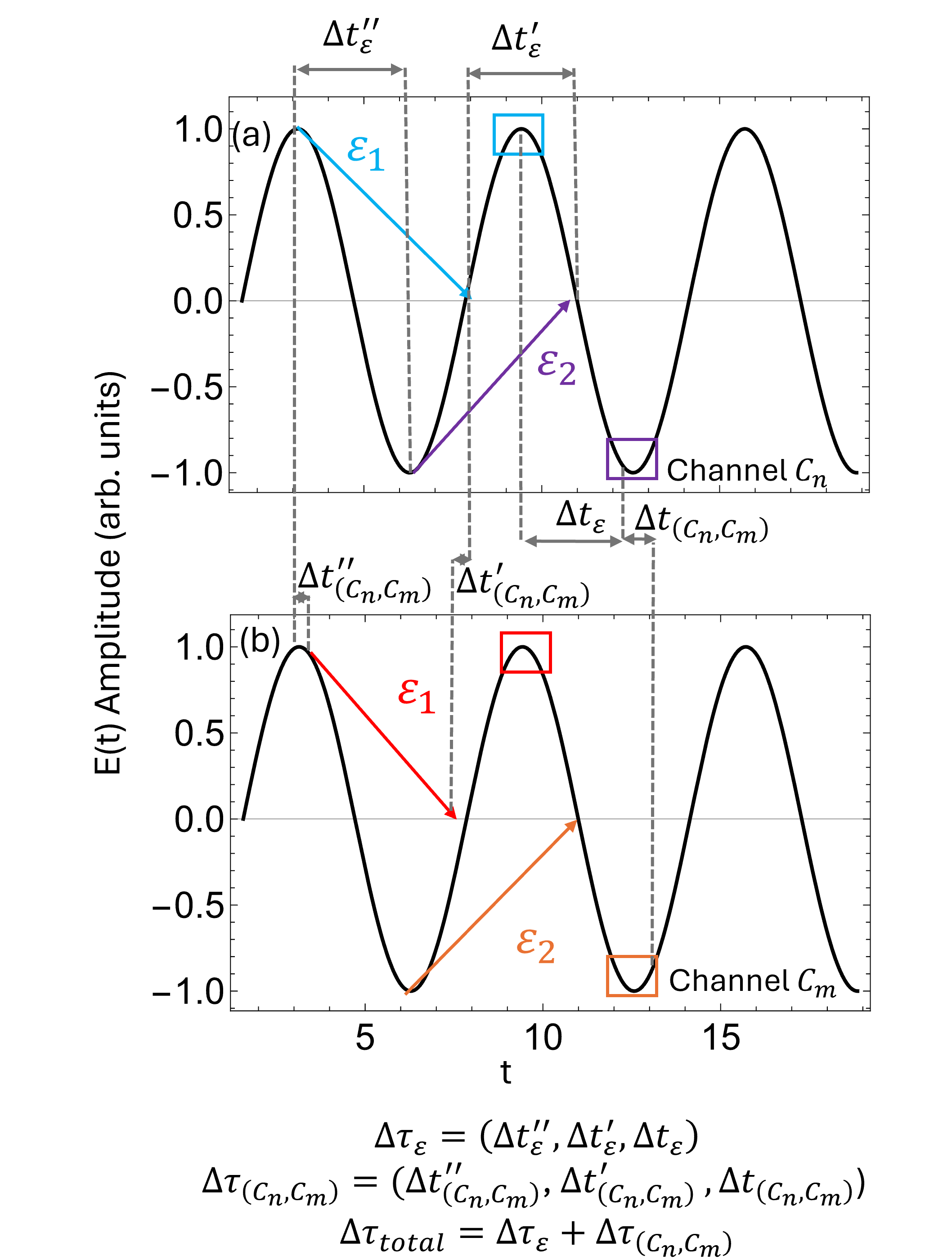}
    \caption{Temporal shifts between (a) events within a single channel, ($\Delta t''_\varepsilon, \Delta t'_\varepsilon, \Delta t_\varepsilon$) and (b) events in different channels ($\Delta t''_{(C_n, C_m)}, \Delta t'_{(C_n, C_m)}, \Delta t_{(C_n, C_m)}$)}
    \label{fig:timeschematic}
\end{figure}

In Fig.~\ref{fig:pulseshape}, we show the $\sin^2$ pulse used in this work. Its length and carrier-envelope phase (CEP) were chosen for continuity with our earlier studies \cite{Faria2012, Hashim2024}, but with reduced peak intensity regarding \cite{Faria2012} to ensure RESI dominance. We adopt the convention of Ref.~\cite{Faria2012}, defining the CEP as $\phi=\phi_1-\phi_0$ with $\phi_0=60^\circ$. In the following figures, $\phi$ denotes $\phi_1$ without the offset. Unless stated otherwise, results are for $\phi=65^\circ$, which we verified against $\phi=155^\circ$. While Ref.~\cite{Hashim2024} analyzed single-channel PMDs and intrachannel interference, here we extend to arbitrary interchannel combinations, which occupy similar momentum regions and exhibit comparable shapes. Thus, our focus is on target-induced interference via excited-state structure, making $\phi=65^\circ$ sufficient for the present purposes.
Figure~\ref{fig:pulseshape} also indicates the relevant events for both electrons, with orbits numbered sequentially from the pulse onset [Fig.~\ref{fig:pulseshape}(a)], and how they are mapped onto the $p_{1\parallel}p_{2\parallel}$ plane [Fig.~\ref{fig:pulseshape}(b)]. We label them $p_io_j$, where $p_i$ denotes the first electron’s ionization–rescattering pair and $o_j$ the second electron’s ionization time \cite{Faria2012, Shaaran2012, Faria1999, Hashim2024}. We focus on the three dominant events marked in Fig.~\ref{fig:pulseshape}, sufficient to probe intra- and intercycle effects. For example, events $p_3o_4$ and $p_4o_5$ have ionization and rescattering times that are approximately all half a cycle apart, whereas $p_5o_6$ occurs one full cycle after $p_3o_4$. More generally, the time displacement between two events may differ for their ionization and rescattering times. Fig.~\ref{fig:timeschematic}(a) shows how we define this time displacement, $\Delta \tau_{(\varepsilon, \varepsilon')} = (\Delta t''_{(\varepsilon, \varepsilon')}, \Delta t'_{(\varepsilon, \varepsilon')}, \Delta t_{(\varepsilon, \varepsilon')})$ for an arbitrary field.

Fig.~\ref{fig:pulseshape}(b) shows the regions of the parallel momentum plane $p_{1\parallel}p_{2\parallel}$ where the PMD associated with each of these three events is localized, with colors matched to those in panel (a). Specifically, $p_3o_4$ maps to the left–down ($\mu=l,d$) quadrant, $p_4o_5$ to the right–up ($\mu=r,u$), and $p_5o_6$, occurring a full cycle after $p_3o_4$, to the left–down quadrant again. For this field, the quadrant labels $(\mu,\nu)$ suffice to identify interfering events. In more complex fields (e.g., bichromatic fields with multiple ionization/rescattering events per half cycle), transition amplitudes must be written as $M_{\mu,\nu}^{(\varepsilon_{1}^{(\mathcal{C}_n)}, \varepsilon_{2}^{(\mathcal{C}_m)})}$ to avoid ambiguity \cite{Hashim2024b}, where $\varepsilon_1, \varepsilon_2$ are the interfering events associated with channels $\mathcal{C}_n$ and $\mathcal{C}_m$. Here, we simplify to $M_{\mu,\nu}^{(\mathcal{C}_{n}, \mathcal{C}_{m})}$ since $\mu,\nu$ uniquely identifies the events in each channel. For events occupying the same momentum regions, but temporally displaced by a full optical cycle, in the remainder of this article we employ the subscript $T$, i.e. $(\mu=l_T,d_T)$, indicating that it belongs to a different optical cycle. 

Figure~\ref{fig:pulseshape}(c) shows the PMD mapping of the same events as in panel (b), but in a different channel. Since the excited-state energy gaps differ across channels, the associated ionization and rescattering times are slightly shifted, despite identical laser parameters. This shift is illustrated in Fig.~\ref{fig:timeschematic}(b), where grey dashed lines mark the interchannel time displacement $\Delta \tau_{(\mathcal{C}_n,\mathcal{C}_m)}$, which may differ for ionization and rescattering. Thus, the total relative temporal displacement between events in different channels can be denoted as $\Delta \tau_{\text{total}} = \Delta \tau_{(\varepsilon, \varepsilon')}  + \Delta \tau_{(\mathcal{C}_m, \mathcal{C}_n)}$.

\section{Generalized interchannel interference conditions}
\label{sec:expinterf}

\subsection{Diagrammatic representation}
\label{sec:diagrammaticinterf}

\begin{figure*}
    \centering
    \includegraphics[width=\textwidth]{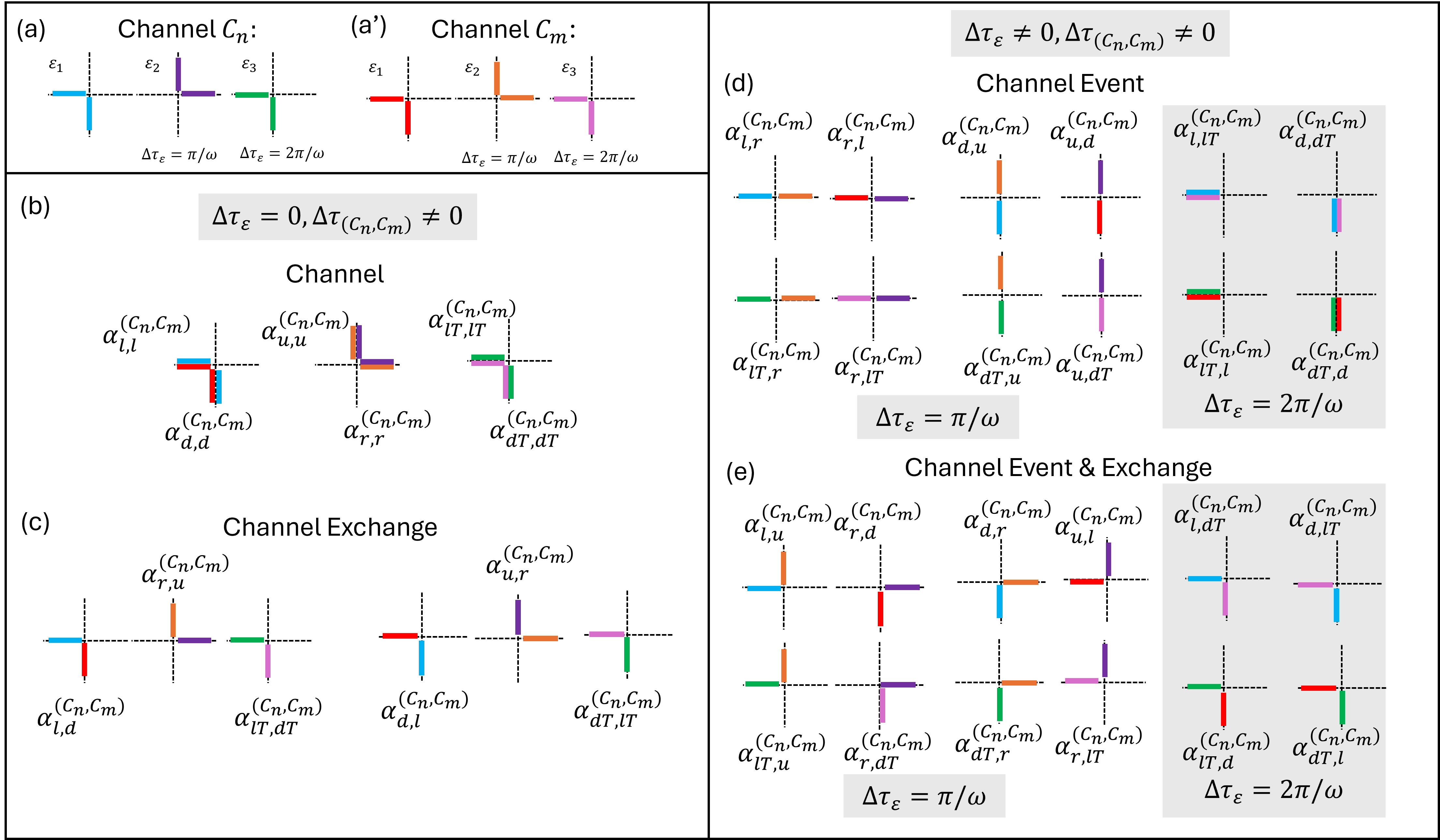}
    \caption{Schematic representation of different types of pair-wise interchannel interference that may occur for NSDI RESI. Panels (a) and (a') show the locations of the transition amplitudes (and actions) for three different events in the parallel momentum plane, for two different channels. $\varepsilon_2$ occurs in the same cycle as $\varepsilon_1$ and will thus exhibit intracycle interference effects, whilst $\varepsilon_3$ occurs in a different field cycle to $\varepsilon_1$ and so interference between these events will be intercycle. Panel (b) shows processes where only channels are summed over coherently and events and symmetrization are incoherent, thus isolating interference occurring from coherent channels. In panel (c), electron momenta are swapped and channels are summed coherently but intracycle events are not. In panel (d), intracycle events and channels are summed coherently but there is no symmetrization. In panel (e), an event and the symmetrized counterpart of another event in the same cycle are summed over coherently, along with the channels. In panels (b)-(e), the phase differences are denoted by $\alpha_{\mu,\nu}$ where $\mu,\nu=l,d,r,u$ and the subscript $T$ denotes that the event occurs in a different field cycle relative to $\varepsilon_1$. We consider only one event per half-cycle here for simplicity, although one can generalize - see \cite{Hashim2025}.}
    \label{fig:schematic2}
\end{figure*}

The generic phase difference associated with interference between two arbitrary events from any two excitation channels is given by $\alpha_{\mu, \nu}^{(\mathcal{C}_{n}, \mathcal{C}_{m})} = S_\mu^{(\mathcal{C}_{n})}-S_\nu^{(\mathcal{C}_{m})}$ where $S_\mu^{(\mathcal{C}_{n})}$ corresponds to a generic action for the n-th channel related to the photoelectron yield in the $\mu$ (= left, down, up or right) part of the $p_{1\parallel}p_{2\parallel}$ plane. Both intracycle and intercycle events may interfere. 

In \cite{Hashim2024}, we identified three types of intrachannel interference, associated with: (1) pure exchange phases from symmetrizing individual events, (2) pure temporal phases from time-displaced events without symmetrization and (3) combined exchange-temporal phases from an event and the symmetrized counterpart of a time-displaced event. Time displacement of events is as defined in Fig.~\ref{fig:timeschematic}(a).

In this work, interchannel interference is investigated in conjunction with these three intrachannel interferences. For simplicity, we focus on two-channel interference, although in principle, one can extend to any number of channels, i.e., it is possible to find pairwise phase differences and add them up. This leads to four types of interchannel interferences. They are illustrated diagrammatically in Fig.~\ref{fig:schematic2} using the event-to-PMD mappings for the two arbitrary channels depicted in Figs.~\ref{fig:schematic2}(a) and (a') (see also the pulse in Fig.~\ref{fig:pulseshape}), but can be extended to an arbitrary number of events in a channel. They are: (1) pure channel phases associated with summing individual unsymmetrized events from different channels [Fig.~\ref{fig:schematic2}(b), computed using process (d) in Table \ref{tab:possiblesums}], (2) channel-exchange phases from summing individual events from one channel with their symmetrized counterparts from a different channel [Fig.~\ref{fig:schematic2}(c), process (b) in Table \ref{tab:possiblesums}], (3) channel-temporal phases from summing an event from one channel with a time-displaced event (intercycle sums highlighted in grey) from a different channel [Fig.~\ref{fig:schematic2}(d), Table \ref{tab:possiblesums}(c)] and (4) channel exchange-temporal phases from an event in one channel and the symmetrized counterpart of a time-displaced event from a different channel (intercycle sums highlighted in grey) [Fig.~\ref{fig:schematic2}(e), computed using Eq.~\eqref{eq:combinedP}]. 

The channel-only interference will occupy the same momentum region as the contributing events i.e. left, down, right or up (i.e. half the momentum axes), since it will involve coherent sums of amplitudes $M_{\mu,\mu}^{(C_n,C_m)} = M_{\mu}^{C_m} + M_{\mu}^{C_n}$, where $\mu=l,d,r,u$ with a subscript $T$ for events displaced by a full cycle. If the same event does not occupy the same regions in different channels due to, e.g., prefactor biases, then this interference will be weak. We denote the corresponding phase difference by $\alpha_{\mu,\mu}^{(\mathcal{C}_{n}, \mathcal{C}_{m})}$. Note that the phases $|\alpha_{\mu,\mu}^{(\mathcal{C}_{n}, \mathcal{C}_{m})}| = |\alpha_{\nu,\nu}^{(\mathcal{C}_{n}, \mathcal{C}_{m})}|$ and are mirror images about the diagonal for $\mu,\nu = l,d$ and $r,u$ (and $l_T,d_T$).

Channel-exchange interference will occupy the first (right-up) and third (left-down) quadrants of parallel momentum plane, since this involves the coherent sums of amplitudes $M_{\mu,\nu}^{(C_n,C_m)} = M_{\mu}^{C_m} + M_{\nu}^{C_n}$ where $\mu,\nu = l,d$ or $r,u$. The channel-exchange phase differences satisfy $|\alpha_{\mu,\nu}^{(\mathcal{C}_{n}, \mathcal{C}_{m})}| = |\alpha_{\nu,\mu}^{(\mathcal{C}_{n}, \mathcal{C}_{m})}|$ for $\mu,\nu = l,d$ and $r,u$ (and $l_T, d_T$). Swapping around the indices $\mu, \nu$ yields phase differences that mirror images about the diagonal. There are twice as many channel-only, and channel-exchange phase differences as there are number of events ($n_\varepsilon$ = 3 here) due to symmetrization. 

Channel-temporal interference occurs between temporally displaced events from different channels. For intracycle displaced events, the resulting interference spans the full momentum axes (left–right or up–down), whereas intercycle displaced events occupy only half of the momentum axes. If temporally displaced events in different channels occupy very different regions, there will be very little interference. 
Interchannel temporal interference is expected to be more prominent than intrachannel temporal interference because this interference type relies on the temporal overlap between the events. Identical events in different channels may have greater overlap than between events within the same channel.

Channel-temporal contributions arise from coherent sums of amplitudes of the type $M_{\mu,\nu}^{(C_n,C_m)} = M_{\mu}^{C_m} + M_{\nu}^{C_n}$ with $\mu,\nu = l,r$; $u,d$ or $l_T,r$ for intracycle interference and $l,l_T$ or $d,d_T$ for intercycle interference. The phase differences appear in transposed pairs, as a consequence from electron symmetrization. For example, $|\alpha_{\mu,\nu}^{(\mathcal{C}_{n}, \mathcal{C}_{m})}| = |\alpha_{\mu',\nu'}^{(\mathcal{C}_{n}, \mathcal{C}_{m})}|$ for $\mu,\nu=l,r$ ($r,l$; $l_T,r$; $r,l_T$; $l,l_T$) and $\mu',\nu'=d,u$ ($u,d$; $d_T,u$; $u,d_T$; $d_T,d$).

Channel–exchange–temporal interference combines both exchange and temporal displacement. For intracycle events, it occupies the second and fourth quadrants of the momentum plane; for intercycle events it appears in the first and third quadrants. They involve the coherent sums of amplitudes $M_{\mu,\nu}^{(C_n,C_m)} = M_{\mu}^{C_m} + M_{\nu}^{C_n}$ where $\mu,\nu = l,u$, $r,d$ or $l_T,u$ for intracycle events and $l,d_T$ or $d,l_T$ for intercycle events. The phase differences $|\alpha_{\mu,\nu}^{(\mathcal{C}_{n}, \mathcal{C}_{m})}| = |\alpha_{\mu',\nu'}^{(\mathcal{C}_{n}, \mathcal{C}_{m})}|$ for $\mu,\nu=l,u$ ($r,d$; $l_T,u$; $r,d_T$; $l,d_T$) and $\mu',\nu'=d,r$ ($u,l$; $d_T,r$; $u,l_T$; $d_T,l$) are equal to each other and mirror images in the diagonal. 
For channel-temporal and channel exchange-temporal interferences, note that the intracycle phase differences $|\alpha_{\mu,\nu}^{(\mathcal{C}_{n}, \mathcal{C}_{m})}| \neq |\alpha_{\nu,\mu}^{(\mathcal{C}_{n}, \mathcal{C}_{m})}|$ for fields that break half-cycle symmetry such as the few-cycle pulse. 

There are $2 (n_\varepsilon^2-n_\varepsilon)$ channel-temporal and channel exchange-temporal phase differences where the factor 2 is due to symmetrization. They can be subdivided into  $2(2n_\varepsilon-2)$ intracycle phase differences and $2(n_\varepsilon^2-3n_\varepsilon-2)$ intercycle phase differences.  Note that we subtract $n_\varepsilon$ `diagonal' cases from the total. This is because channel-only interference is a special case of channel-temporal interference where there is no relative time displacement of the events $\Delta \tau_{(\varepsilon, \varepsilon}) = 0$. Likewise, channel-exchange interference is a special case of channel temporal-exchange interference where the relative time event time displacement vanishes. 

For the pulse, we consider three events as shown in Fig.~\ref{fig:pulseshape}, analogous to the $\varepsilon_1, \varepsilon_2, \varepsilon_3$ events in Fig.~\ref{fig:schematic2}(a). For two-channel interference, one can therefore compute 6 channel-only phase differences, 6 channel-exchange interferences, 12 channel-temporal interferences and 12 channel-combined interferences all of which are depicted schematically in Fig.~\ref{fig:schematic2}(b)-(e). 

\subsection{Analytic expressions}

In all derivations that follow, we consider interference between two channels $n$ and $m$, where the ionization, rescattering and tunneling times of the $n$-th channel are given by $t''_n, t'_n$ and $t_n$. We take the times of the $m$-th channel to $t''_m = t''_n+\Delta t''_{(n,m)}, t'_m = t'_n+\Delta t'_{(n,m)}$ and $t_m = t_n+\Delta t_{(n,m)}$ where $\Delta \tau_{(n,m)} = (\Delta t''_{(n,m)}, \Delta t'_{(n,m)}, \Delta t_{(n,m)})$ are the difference in times, for an identical event, between the n and m-th channels. This is shown schematically in Fig.~\ref{fig:timeschematic}. For energetically close channels, $\Delta \tau_{(n,m)} \rightarrow 0$. Note that we do not extend these interference derivations to N-channel sums because of the increasing complexity of the phase differences with more than two channels. Even with two channels, there will be ${N\choose 2}$ overlapping (pairwise) phase differences. As previously stated, we neglect multiple ionization events per half cycle, although the following can be generalized for such events using a different time displacement - see \cite{Hashim2025} for intrachannel conditions where we show how to account for such events explicitly.

We first derive analytic conditions for channel-only interference [Fig.~\ref{fig:schematic2}(b)], yielding phase differences with temporal building blocks analogous to intrachannel interference. For an event mapping to the third quadrant of the $p_{1\parallel}p_{2\parallel}$ plane, the channel-only phase is given by\footnote{Note that this channel-only phase difference, $\alpha_{l,l}^{(\mathcal{C}_n, \mathcal{C}_m)}$, differs from the $\alpha_{l,l}^{(m,n)}$ phase difference employed in \cite{Hashim2025} where $m, n$ refer to different events within a channel, whereas this work neglects multiple ionization events per half cycle for simplicity but this convention can be extended in principle.}

\begin{eqnarray}
    \alpha_{l,l}^{(\mathcal{C}_n, \mathcal{C}_m)} &=& S_l^{(\mathcal{C}_n)} - S_l^{(\mathcal{C}_m)} 
    \nonumber\\ &=& -\alpha^{(\text{ene})}_{\Delta \tau_{(n,m)}} + \alpha^{(\text{shift})}(t_n, t'_n) \nonumber\\ &+& \frac{1}{2}\alpha^{(\text{pond})}_{\Delta \tau_{(n,m)}}(t''_n, t_n) - \alpha^{(A^2)}_{\Delta \tau_{(n,m)}}(t''_n, t'_n) \nonumber\\ &-& \alpha^{(\mathbf{p}_1, \mathbf{p}_2)}_{\Delta \tau_{(n,m)}}(t'_n,t_n)
    \label{eq:channelonly}
\end{eqnarray}
with $\Delta \tau = \Delta \tau_{(n,m)}$, 
where
\begin{eqnarray}
    \alpha^{(\text{ene})}_{\Delta \tau_{(n,m)}} &=& E_{1g}^{(\mathcal{C}_m)} \Delta t''_{(n,m)} + 
    E_{2e}^{(\mathcal{C}_m)} \Delta t_{(n,m)} \nonumber\\ &+& \Delta t'_{(n,m)}(E_{2g}^{(\mathcal{C}_n)} - E_{2e}^{(\mathcal{C}_m)}) \nonumber\\ &+& \frac{\mathbf{p}_1^2\Delta t'_{(n,m)}}{2} + \frac{\mathbf{p}_2^2\Delta t_{(n,m)}}{2},
    \label{eq:alphaene}
\end{eqnarray}

\begin{eqnarray}
    \alpha^{(\mathbf{p}_1,\mathbf{p}_2)}_{\Delta \tau_{(n,m)}}(t'_n, t_n) &=& \mathbf{p}_1 \cdot [\mathbf{F}_{A}(t_n' + \Delta t'_{(n,m)}) - \mathbf{F}_{A}(t_n')] \nonumber \\ &+& \mathbf{p}_2 \cdot [
     \mathbf{F}_{A}(t_n + \Delta t_{(n,m)}) - \mathbf{F}_{A}(t_n)  ],
    \label{eq:alphap1p2}
\end{eqnarray}
\begin{eqnarray}
    \alpha^{(\text{pond})}_{\Delta \tau_{(n,m)}}(t''_n, t'_n) &=& F_{A^2}(t_n'' + \Delta t''_{(n,m)}) + F_{A^2}(t_n + \Delta t_{(n,m)}) \nonumber\\ &-& F_{A^2}(t_n) -  F_{A^2}(t_n'') 
    \label{eq:alphapond}
\end{eqnarray}
and 
\begin{eqnarray}
    \alpha^{(A^2)}_{\Delta \tau_{(n,m)}}(t''_n, t'_n) &=& \frac{[\mathbf{F}_{A}(t_n' + \Delta t'_{(n,m)})-\mathbf{F}_{A}(t''_n + \Delta t''_{(n,m)})]^2}{2(t'_n-t''_n+\Delta t'_{(n,m)}-\Delta t''_{(n,m)})} \nonumber\\ &-& \frac{[\mathbf{F}_{A}(t_n')- \mathbf{F}_{A}(t''_n)]^2}{2(t'_n-t''_n)}
    \label{eq:alphaA2}
\end{eqnarray}
are the temporal shift building blocks as derived in \cite{Hashim2024, Hashim2024b}, with 
\begin{equation}
    \mathbf{F}_A(t)=\int^t \mathbf{A}(\tau)d\tau,
    \label{eq:IntegralA}
\end{equation}
\begin{equation}
    F_{A^2}(t)=\int^t \mathbf{A}^2(\tau)d\tau.
    \label{eq:IntegralA2}
\end{equation}

There is an additional phase difference proportional to the energy difference of the excited states and the time delay between rescattering and tunneling,

\begin{eqnarray}
    \alpha^{(\text{shift})}(t_n, t'_n) &=& (t_n-t'_n)(E_{2e}^{(\mathcal{C}_{n})} - E_{2e}^{(\mathcal{C}_{m})}).
    \label{eq:alphashift}
\end{eqnarray}
The energy-dependent term introduces corrections proportional to the channel energy-gap difference and $(t-t')$. The simplest case for constructive interference $n=0$ requires each building block to vanish as studied previously for intrachannel interference \cite{Maxwell2015,Hashim2024,Hashim2025}. However, the interchannel energy shift never vanishes since neither the energy gap difference nor $(t'-t)$ is zero. Thus, only $n\neq0$ is relevant, yielding a nonzero phase. Constructive interference occurs only when this phase is an even multiple of $\pi$; otherwise, it mainly shifts the fringes.

The building blocks produce distinct interference contributions. $\alpha^{\text{ene}}_{\Delta \tau}$ describes a hypersphere and yields circular fringes for $\Delta t''=\Delta t'=\Delta t$ and large $\Delta \tau$, but these conditions are not met, so its effect is minimal. $\alpha^{(A^2)}_{\Delta \tau}$ depends only on the first electron’s ionization and return times, shaped by $\mathbf{F}_A(\tau)$; it induces small shifts and may form linear alternating fringes, which upon incoherent symmetrization yield chequerboard-like PMD patterns. The ponderomotive block $\alpha^{(\text{pond})}_{\Delta \tau}$ involves ${F}_{A^{2}}(t)$ [Eq.~\eqref{eq:IntegralA2}], a smooth monotonic function causing numerical shifts without altering interference shape. $\alpha^{(\mathbf{p}_1,\mathbf{p}_2)}_{\Delta \tau}$ produces diagonal fringes, with spacing and intercept set by $\mathbf{F}_A(t)$ [Eq.~\eqref{eq:IntegralA}]. Channel-only interference thus combines all these contributions, but its interplay is hard to predict. In intrachannel PMDs, rich structures (wings, chequerboards, v-shaped fringes, circular substructures) were observed \cite{Hashim2024,Hashim2025}, whereas channel-only interference should be simpler: it requires events from both channels to occupy the same region, which depends strongly on state geometry. Moreover, interference  contrast is suppressed if one channel dominates, and state energies influence both the PMD localization and the size of $\Delta \tau_{(n,m)}$ and $\alpha^{\text{(shift)}}$.

The second type of two-channel interference is channel-exchange, which occurs between an event in one excitation channel and its symmetrized counterpart in another channel as depicted in Fig.~\ref{fig:schematic2}(c). 
For an event mapping to the third quadrant of the $p_{1\parallel}p_{2\parallel}$ plane, the corresponding phase difference is 
\begin{eqnarray}
    \alpha_{l,d}^{(\mathcal{C}_n, \mathcal{C}_m)} &=& S_l^{(\mathcal{C}_n)} - S_d^{(\mathcal{C}_m)} 
    \nonumber\\ &=& -\alpha^{(\text{ene}, \mathbf{p}_1 \leftrightarrow \mathbf{p}_2)}_{\Delta \tau_{(n,m)}} + \frac{1}{2}\alpha^{(\text{pond})}_{\Delta \tau_{(n,m)}}(t''_n, t_n) \nonumber\\&- &\alpha^{(A^2)}_{\Delta \tau_{(n,m)}}(t''_n, t'_n) -\alpha^{(\mathbf{p_2 \leftrightarrow p_1})}_{\Delta \tau_{(n,m)}}(t'_n,t_n)\nonumber \\ &-& \alpha_{(\mathbf{p}_1, \mathbf{p}_2)}^{(\text{exch})}(t'_n,t_n),
    \label{eq:channelexchange}
\end{eqnarray}
where
\begin{eqnarray}
    \alpha_{(\mathbf{p_1, p_2})}^{(\text{exch})}(t'_n,t_n) &=& \frac{1}{2} (\mathbf{p}_2^2 - \mathbf{p}_1^2)(t_n - t'_n)
    \label{eq:alphaexch}
\end{eqnarray}
and
\begin{eqnarray}
   \hspace*{-0.15cm} \alpha^{(\mathbf{p}_2 \leftrightarrow \mathbf{p}_1)}_{\Delta \tau_{(n,m)}}(t'_n,t_n) &=& \hspace*{-0.08cm}\mathbf{p}_1 \cdot [\mathbf{F}_{A}(t_n + \Delta t_{(n,m)}) - \mathbf{F}_{A}(t_n')] \nonumber \\ &+&\hspace*{-0.08cm} \mathbf{p}_2 \cdot [
     \mathbf{F}_{A}(t'_n + \Delta t'_{(n,m)}) - \mathbf{F}_{A}(t_n)  ]
    \label{eq:alphatempexch}
\end{eqnarray}
are the building blocks associated with exchange, and temporal-exchange respectively. The form of this channel-exchange phase difference is expected due to the condition $\tau_m = \tau_n + \Delta \tau_{(n,m)}$. Note that the momenta in Eq.~\eqref{eq:channelexchange} are swapped, denoted by $\mathbf{p}_1\leftrightarrow\mathbf{p}_2$ in the superscript.

Beyond the energy, ponderomotive, and $A^2$ terms, the field-independent phase $\alpha^{(\mathrm{exch})}_{\mathbf{p}_1,\mathbf{p}_2}(t,t')$ generates hyperbolae in the $(p_{1\parallel},p_{2\parallel})$ plane with asymptotes $p_{1\parallel}=\pm p_{2\parallel}$ and eccentricity $\sqrt{2}$. For linearly polarized fields, channel exchange also produces diagonal or antidiagonal fringes, including a possible central “spine.” Temporal blocks further induce phase jumps, $v$-shaped fringes, and faint chequerboard-like patterns  \cite{Hashim2024}. Overall, channel-exchange interference is more complex than intrachannel interference and occupies the full overlap of the single-channel PMDs.

An event occurring roughly half a cycle before or after yields probability densities and amplitudes localized in the first quadrant, associated with phase difference $\alpha_{ru}$ gives
\begin{eqnarray}
    \alpha_{r,u}^{(\mathcal{C}_n, \mathcal{C}_m)} &=& S_r^{(\mathcal{C}_n)} - S_u^{(\mathcal{C}_m)} 
    \nonumber\\ &=& -\alpha^{(\text{ene})}_{\Delta \tau_{\text{total}}} + \frac{1}{2}\alpha^{(\text{pond})}_{\Delta \tau_{\text{total}}}(t''_n, t_n) \nonumber\\ &-&  \alpha^{(A^2)}_{\Delta \tau_{\text{total}}}(t''_n, t'_n) - \alpha^{(\mathbf{p}_2 \leftrightarrow \mathbf{p}_1)}_{\Delta \tau_{\text{total}}}(t'_n,t_n)\nonumber\\ &-& \alpha_{(\mathbf{p}_1, \mathbf{p}_2)}^{(\text{exch})}(t'_n + \Delta \tau_\varepsilon,t_n + \Delta \tau_\varepsilon)
    \label{eq:channelexchange2}
\end{eqnarray}
where $\Delta \tau_{\text{total}} = \Delta\tau_{(n,m)}+ \Delta \tau_\varepsilon$ is the total time displacement. $\Delta \tau_\varepsilon$ describes the time-displacement relative to the event occupying the left-down quadrant and is roughly $\pi/\omega$ for intracycle events and $2\pi/\omega$ for intercycle events. Due to the lack of half cycle symmetry for the pulse,  $\alpha_{ru} \neq \alpha_{ld}$.

The third interference type is channel-event (or channel-temporal) interference [Fig.~\ref{fig:schematic2}(d)], arising from temporally shifted events in different channels with incoherent symmetrization [process (c), Table~\ref{tab:possiblesums}]. Compared to intrachannel interference, the additional $\Delta \tau_{(n,m)}$ increases the total time displacement.  This is expected to enhance the effect of some temporal building block terms. 

Therefore,
\begin{eqnarray}
    \alpha_{l,r}^{(\mathcal{C}_n, \mathcal{C}_m)} &=& S_l^{(\mathcal{C}_n)} - S_r^{(\mathcal{C}_m)} 
    \nonumber\\ &=& -\alpha^{(\text{ene})}_{\Delta \tau_{\text{total}}} + \alpha^{(\text{shift})}(t_n, t'_n) \nonumber\\ &+& \frac{1}{2}\alpha^{(\text{pond})}_{\Delta \tau_{\text{total}}}(t''_n, t_n) - \alpha^{(A^2)}_{\Delta \tau_{\text{total}}}(t''_n, t'_n) \nonumber\\ &-& \alpha^{(\mathbf{p}_1, \mathbf{p}_2)}_{\Delta \tau_{\text{total}}}(t'_n,t_n),
    \label{eq:channeltemporal}
\end{eqnarray}
where the building blocks are as given in Eqs.~\eqref{eq:alphaA2}-~\eqref{eq:alphashift}. This interference yields the same shapes as channel-only interference but is weaker, since it requires overlapping events from different channels without one dominating. Moreover, $\Delta \tau_\varepsilon$ depends on field symmetry \cite{Rook2022}.

The fourth and final interference type is channel temporal-exchange depicted in Fig.~\ref{fig:schematic2}(e). In this process, an event in one channel and the symmetrized counterpart of a time-displaced event from a different channel are summed coherently. The phase difference involving events in the first and third quadrant is given by
\begin{eqnarray}
    \alpha_{l,u}^{(\mathcal{C}_n, \mathcal{C}_m)} &=& S_l^{(\mathcal{C}_n)} - S_u^{(\mathcal{C}_m)} 
    \nonumber\\ &=& -\alpha^{(\text{ene}, \mathbf{p_1 \leftrightarrow p_2})}_{\Delta \tau_{\text{total}}} + \frac{1}{2}\alpha^{(\text{pond})}_{\Delta \tau_{\text{total}}}(t''_n, t_n) \nonumber\\ &-&\alpha^{(A^2)}_{\Delta \tau_{\text{total}}}(t''_n, t'_n) \nonumber-\alpha^{(\mathbf{p}_2 \leftrightarrow \mathbf{p}_1)}_{\Delta \tau_{\text{total}}}(t'_n,t_n)\\ &-& \alpha_{(\mathbf{p}_1, \mathbf{p}_2)}^{(\text{exch})}(t'_n,t_n).
    \label{eq:channelcombined}
\end{eqnarray}
This will lead to the same interference patterns as channel-exchange interference, with changes in spacings, intercepts and gradients, perhaps leading to closer fringes forming `fishbone' like structures in the second and fourth quadrants, near the axes where the events may overlap. The other building blocks are expected to play only a minimal role. Note the lack of the $\alpha^{\text(shift)}$ building block. Given its lack of dependence on the energy shift, this type of interference is not expected to be particularly prominent for any two-channel combination.

\section{Correlated electron momentum distributions}
\label{sec:momdists}
\begin{figure}
    \centering
    \includegraphics[width=\columnwidth]{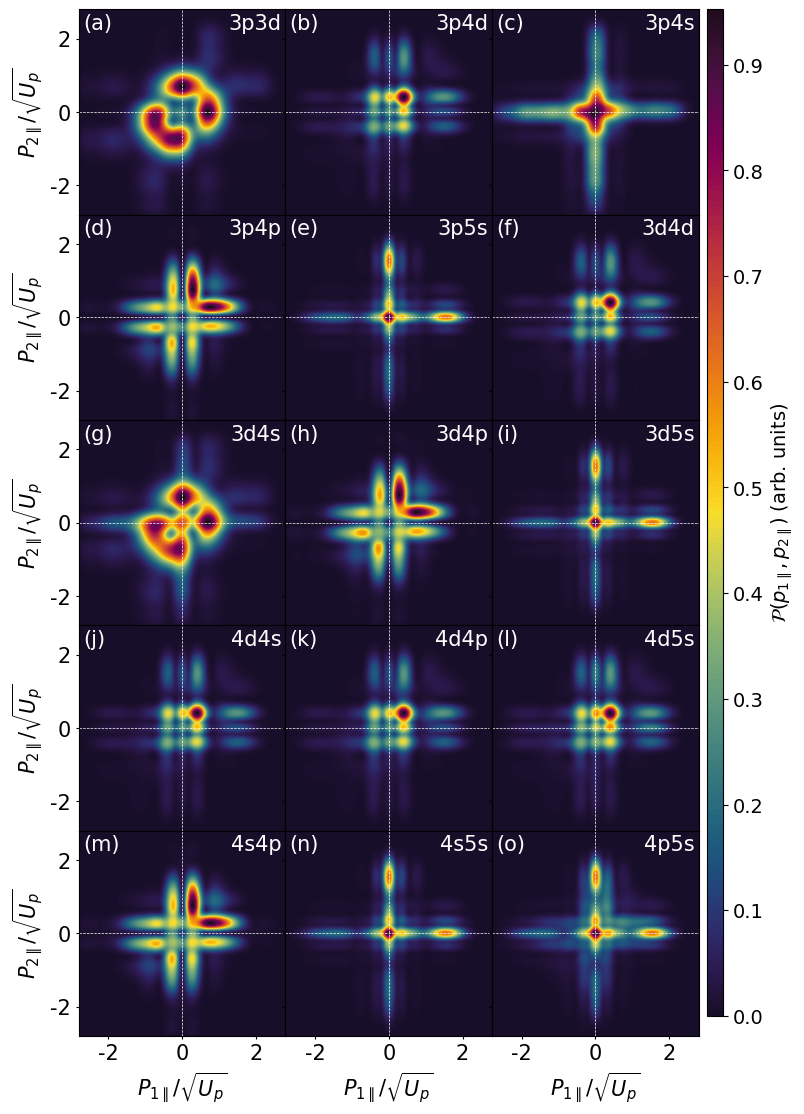}
    \caption{Fully incoherent two-channel momentum distributions for all channels in Table \ref{tab:channels}, $\mathcal{P}_{iii} (p_{1\parallel}, p_{2\parallel})$ [given by Table ~\ref{tab:possiblesums}(h)] computed for Argon with a linearly polarized few-cycle pulse with the same parameters as in Fig.~\ref{fig:pulseshape} and taking the three most dominant events. The axes are indicated with white dashed lines. The excited states for each pair of channels is shown in the top-right corner. For example, $3p4p$ indicates that the $3s\rightarrow3p$ and $3p\rightarrow4p$ transitions have been combined.}
    \label{fig:incoherent}
\end{figure}

We begin by examining incoherent two-channel PMDs to identify cases where channel contributions are comparable. Among the fifteen pairwise sums shown in Fig.~\ref{fig:incoherent}, the two PMDs with the smallest excitation energy difference ($E_\text{diff}=0.01$ a.u.), namely $3d4s$ and $4d5s$ [Fig.~\ref{fig:incoherent}(c) and (l)], deviate most noticeably from the characteristic structures associated with pure $s$, $p$, or $d$ states, instead appearing as hybrid mixtures of $d$- and $s$-state features. 

In nearly all other two-channel PMDs, one channel dominate and the PMD contains the characteristic features, summarized in the final column of Table~\ref{tab:channels}: maxima along the axes (a cross, indicative of $s$ states), along the diagonals ($p$ states), or in both regions ($d$ states). For reference, the single-channel PMDs for all six channels are shown in Fig.~\ref{fig:singlechannel} in the Appendix, with a detailed discussion in \cite{Hashim2024}.
In particular, $4s5s$, $3d5s$, $3p5s$, and $3p4s$ [Fig.~\ref{fig:incoherent}(n),(i),(e),(c)] are dominated by $s$ states; $3p4p$, $3d4p$, $4s4p$, and $3p3d$ [Fig.~\ref{fig:incoherent}(d),(h),(m),(a)] by $p$ states; and $3p4d$, $3d4d$, and $4d4p$ [Fig.~\ref{fig:incoherent}(b),(f),(k)] by $d$ states.
In several cases, such as $4s5s$, $3p4p$, and $3d4d$ [Fig.~\ref{fig:incoherent}(n),(d) and (f)], this behavior is anticipated given the contributing states have the same orbital angular momentum.  Thus, the bound-state geometry embedded in the prefactor $V_{\mathbf{p}_2e}$ is mapped similarly to the $p_{1\parallel}p_{2\parallel}$ plane. The $4p5s$ case [Fig.~\ref{fig:incoherent}(o)], despite its moderate energy difference (0.12) also appears to have comparable contributions from the two channels. This is particularly evident when contrasted with $3p5s$ [Fig.~\ref{fig:incoherent}(e)], there is a significant enhancement of secondary maxima in the first quadrant, at the locations where the $3p$ PMD is maximal.

This raises the question, under what conditions does one channel dominate the incoherent PMD? As outlined in the background, the factors influencing the PMDs include the relative intensities of the single-channel PMDs, the spatial geometry of the excited states and the excitation energy gap between the channels which governs the dynamics of the process and leads to competing temporal effects. If the excitation events in the two channels are mapped into non-overlapping momentum regions, the resulting incoherent sum may manifest as a broader distribution—more representative of a mixture than an interference pattern. In such scenarios, the likelihood of interference is diminished. We next examine how each factor shapes the incoherent PMDs, and then assess their impact on interference via the coherent two-channel PMDs.

\subsection{Estimating comparable channels}
\label{sec:comparablechannels}

\begin{table}[h]
\centering
\begin{adjustbox}{max width=280pt}
\begin{tabular}{cccccc}
\hline \hline
Excited States & $E_\text{diff}$ & $M_\text{diff}$ & $O_R$ & EMM \\
\hline \hline
\textbf{3p3d} & 0.11 & 1 & -0.225 & 0.380\\
\textbf{3p4d} & 0.34 & 4 & 0.426 & 0.053\\
3p4s & 0.12 & 1 & -0.229 & 0.550\\
3p4p & 0.21 & 3 & 0.289 & 0.179\\
3p5s & 0.33 & 3 & -0.095 & 0.084\\
3d4d & 0.23 & 3 & 0.327 & 0.150\\
\textbf{3d4s} & 0.01 & 0 & -0.073 & 0.747\\
3d4p & 0.10 & 2 & 0.041 & 0.518\\
3d5s & 0.22 & 2 & -0.088 & 0.353\\
4d4s & 0.22 & 3 & 0.249 & 0.103\\
4d4p & 0.13 & 1 & -0.291 & 0.383\\
\textbf{4d5s} & 0.01 & 1 & -0.085 & 0.549\\
4s4p & 0.09 & 2 & -0.983 & 0.346\\
\textbf{4s5s} & 0.21 & 2 & 0.293 & 0.211\\
\textbf{4p5s} & 0.12 & 0 & -0.249 & 0.771 \vspace*{0.1cm} \\ \hline \hline
\end{tabular}
\end{adjustbox}
\caption{Summary of metrics for all pairwise combinations of excitation channels. The first column lists the channel pairs, with the six cases investigated in detail highlighted in bold. The second column gives the excitation energy gap difference $E_{\text{diff}}$, between the two channels. The third column lists the relative intensity difference of the single-channel PMDs $M_\text{diff}$. The fourth column contains the normalized radial overlap between the excited-state orbitals $O_R$, quantifying their geometric similarity. The final column lists the Equal Mix Metric (EMM), indicating the degree to which the two-channel PMD represents an equal contribution from both channels ($\text{EMM}=1$) or is dominated by one channel ($\text{EMM}=0$).
\label{tab:metrics}} 
\end{table}

To determine which channels dominate and to predict the shape and contrast of two-channel interference, we quantify the influence of three factors. The first is the energy-gap difference, $E^{(\mathcal{C}_n,\mathcal{C}_m)}_{\mathrm{diff}}=\Delta E^{(\mathcal{C}_n)}-\Delta E^{(\mathcal{C}_m)}$ as introduced in Sec.~\ref{sec:target} calculated for each pair of excitation channels - see the second column of Table~\ref{tab:metrics}.

The second factor is the relative intensity of each channel, denoted $M_\text{diff}$ and is given for all channel pairs in the third column of Table \ref{tab:metrics}. This is estimated by computing the average magnitude of the single-channel PMDs and then evaluating the order-of-magnitude difference between the two channels. This serves as a heuristic to identify which channels yield stronger PMDs and are therefore more likely to dominate, potentially suppressing contributions from the weaker channel in the incoherent sum.

The third factor is the geometric similarity of the bound excited states in the two contributing channels. To quantify this, we focus on the radial electron distributions in Argon and compute normalized radial overlap integrals
\begin{equation}
\mathcal{O}_R = \frac{\int^\infty_0 R_{n_1,l_1}(r) R_{n_2,l_2}(r) r^2 dr}{\sqrt{\int^{\infty}_0 |R_{n_1, l_1}(r)|^2 r^2  dr \int^{\infty}_0 |R_{n_2, l_2} (r)|^2 r^2 dr}}
\label{eq:overlap}
\end{equation}
where $R_{n,l}$ are hydrogenic radial wavefunctions corresponding to different channels with effective nuclear charges estimated using Slater’s rules\cite{Clementi1963}. The angular components of the orbitals are omitted to avoid vanishing overlaps arising from spherical-harmonic orthogonality. Note that because the effective nuclear charges vary between different excited-state configurations, the overlap integral will not vanish even between excited states sharing the same angular momentum. 

This procedure yields a symmetric matrix of radial overlaps between orbitals, quantifying their structural similarity and relative localization as a function of $r$. The normalized integral serves as a similarity metric: absolute values near 1 indicate strong radial resemblance whilst values near 0 indicate distinct shapes or localization. The sign can be negative or positive. A negative value indicates that the radial wavefunctions have opposite sign and are therefore out of phase, but does not imply dissimilarity. This measure enables us to predict the extent to which two-channel PMDs may occupy overlapping or distinct regions in the parallel momentum plane.  This quantity has been postulated based on the ionization prefactor $V_{\mathbf{p}_2e}$ associated with the ionization of the second electron being formally similar to the Fourier transform of the second electron's excited state, apart from the term $V_\mathrm{ion}$ in the integrand. 

To directly assess the relative influence of each constituent channel on the resulting two-channel PMD, we introduce the `Equal Mix Metric' (EMM), an influence metric based on the 2D Earth Mover’s Distance (EMD). The EMD is a statistical measure that calculates the minimum ''cost" to transform one distribution into another, taking into account the spatial arrangement of the distributions. This makes it particularly well-suited for comparing PMDs, where spatial structure carries important physical information. Compared to alternative metrics, the EMD more accurately reflects perceptual dissimilarity between distributions. For more details see \cite{Rubner2000}. The EMM is defined as
\begin{eqnarray}
    \text{EMM} = 1-2|\chi_{n}-0.5|
\label{eq:EMM}
\end{eqnarray}
where

\begin{eqnarray}
\chi_{n} = 1 - & \frac{\text{EMD}(\mathcal{P}^{(\mathcal{C}_n)}, \mathcal{P}^{(\mathcal{C}_{n}, \mathcal{C}_m)})}{\text{EMD}(\mathcal{P}^{(\mathcal{C}_n)}, \mathcal{P}^{(\mathcal{C}_{n}, \mathcal{C}_m)}) + \text{EMD}(\mathcal{P}^{(\mathcal{C}_m)}, \mathcal{P}^{(\mathcal{C}_{n}, \mathcal{C}_m)})}
\label{eq:eqmm}
\end{eqnarray}
and $\text{EMD}(\mathcal{P}^{(\mathcal{C}_n)}, \mathcal{P}^{(\mathcal{C}_{n}, \mathcal{C}_m)})$ represents the EMD between each single-channel PMD and the two-channel PMD.
While the EMD itself is well-established \cite{Rubner2000} and has uses in many fields \cite{Orlova2016, Kranstauber2017, Zhang2020emd, Wang2023emd, Komiske2019}, the EMM represents a novel application in this context, providing an intuitive scale for interpreting channel dominance. Appendix \ref{Sec:EMDappendix} provides further details on the specific computation details.

A value of $\text{EMM}=1$ indicates that both channels contribute equally to the two-channel PMD, while $\text{EMM}=0$ indicates that one channel dominates and the resulting PMD closely resembles the dominant channel. In this case, most interference arises from intrachannel processes rather than interchannel effects. Conversely, when $\text{EMM}\rightarrow1$, the PMD exhibits characteristic features from both contributing channels, and interchannel interference fringes are expected to appear prominently, though may be blurred due to the overlapping contributions. Other values of EMM indicate mixing to varying degrees, and it is difficult to predict the shape and contrast of such interference. 

To gain insight into how each factor influences the two-channel PMDs, we compute the Pearson correlation coefficients (PCC) \cite{Asuero2007} between the EMM and each of $E_\text{diff}, M_\text{diff}$ and $O_R$ values listed in Table.~\ref{tab:metrics}. Further details on the PCC can be found in Appendix \ref{Sec:pccAppendix}. The resulting values $\mathcal{R}^{(\text{EMM,} E_{\mathrm{diff}})}=-0.84$, $\mathcal{R}^{(\text{EMM,} M_{\mathrm{diff}})}=-0.9$ and $\mathcal{R}^{(\text{EMM,} |\mathcal{O}_R|)}=0.45$ respectively. These results indicate that the excitation energy difference and relative magnitude are the primary factors determining the degree of mixing between channels, while radial overlap plays a secondary role. Negative (positive) correlation values for $\mathcal{R}^{(\text{EMM, E}_{\mathrm{diff}})}$ and $\mathcal{R}^{(\text{EMM, M}_{\mathrm{diff}})}$ ($\mathcal{R}^{(\text{EMM}, |\mathcal{O}_R|)}$) show that smaller energy and magnitude differences (larger overlap) are associated with more balanced channel contributions as measured by the EMM, as expected. 

From Table~\ref{tab:metrics}, only the $3d4s$ and $4p5s$ pairs exhibit significant equal mixing (EMM $\geq 0.7$). 
These pairs have ideal conditions for equal mixing, given that both have zero magnitude difference, the primary factor determining the degree of mixing. The secondary factors may compete and compensate each other with the energy gap playing a larger role as shown by its PCC value discussed above. For example, the $3d4s$ pair has a minimal energy gap (0.01), offsetting the effect of the least influential radial similarity factor, for which it has a very small value (-0.073). Meanwhile, the  $4p5s$ pair has a moderate energy gap (0.12) but also moderate radial overlap (-0.249) creating conditions for which one obtains a high EMM.
These two pairs are therefore expected to show strong interchannel interference with features arising from both contributing channels.
Most channel pairs with large magnitude differences (3–4 orders of magnitude) have low EMM values, indicating strong suppression of mixing: when one channel dominates in intensity, the weaker channel’s contribution is effectively obscured, regardless of energy or overlap. The most extreme case is $3p4d$, with an EMM of 0.053, making it the least likely to display interchannel interference.

Several intermediate EMM cases are also of interest. For instance, $3p3d$ has a moderate energy gap and radial overlap but low EMM due to its small magnitude; $4d5s$ shows relatively high EMM despite a large magnitude difference; and $4s5s$ exhibits moderate EMM despite poor radial overlap. These cases illustrate how the factors governing mixing can compete and suggest that some interference types may be more sensitive to certain factors than others.
One should note that the EMM is not a factor based on the underlying physics, rather it is a distance between probability distributions and thus involves a degree of arbitrariness.

\subsection{Quantum interference}

\begin{figure}
    \centering
    \includegraphics[width=\columnwidth]{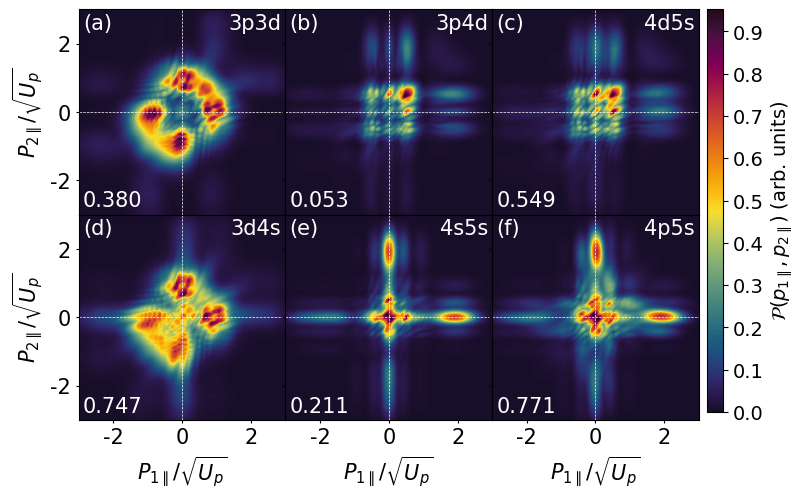}
    \caption{Fully coherent two-channel momentum distributions 
    $\mathcal{P}_{ccc} (p_{1\parallel}, p_{2\parallel})$ [Eq.~\eqref{eq:fullcoherent}]  for select channels in Table \ref{tab:channels}. The axes are indicated with white dashed lines. The excited states for each pair of channels is shown in the top-right corner. All other parameters are the same as in Fig.~\ref{fig:pulseshape}.}
    \label{fig:coherent}
\end{figure}

\begin{figure}
    \centering
    \includegraphics[width=\columnwidth]{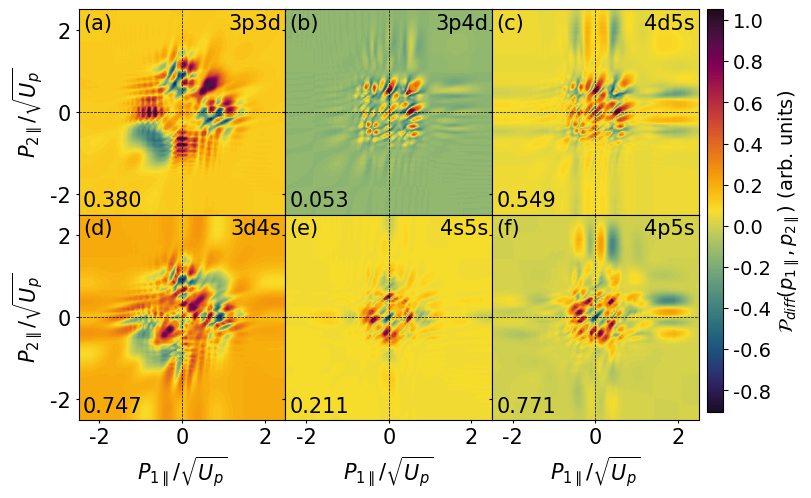}
    \caption{Difference between the fully coherent and fully incoherent RESI distributions $\mathcal{P}_{ccc} (p_{1\parallel}, p_{2\parallel}) -\mathcal{P}_{iii} (p_{1\parallel}, p_{2\parallel})$ for select pairs of channels, computed with the same driving field parameters as in Fig.~\ref{fig:pulseshape}. The excited states for each pair of channels is shown in the top-right corner. The axes are indicated with black dashed lines. The equal mix metric value is indicated in the bottom-left corner of each panel. The signal in each panel has been normalized with regard to its maximum. }
    \label{fig:difference}
\end{figure}

In this section, we present the coherent two-channel PMDs, and systematically disentangle the contributions from the four interference types. We relate features of observable interference in the $p_{1\parallel}p_{2\parallel}$ plane to target and channel properties such as excited-state energy differences, channel intensities, radial overlap of the excited states, as well as the overlap in the mapping of the contributing events for each channel onto the parallel momentum plane. This has implications for allowing for for better prediction and interpretation of interference patterns. 
As established in Sec.~\ref{sec:expinterf}, the interference patterns arising from channel-exchange, channel-temporal, and channel-combined phase differences are expected to have shapes similar to those for the intrachannel case, but due to the number of processes, these patterns will be superimposed on and may blur each other. 

Our analysis is grounded in the Equal Mixing Metric (EMM), which quantifies how evenly a two-channel PMD combines its constituent single-channel PMDs—a concept previously explored only qualitatively \cite{Maxwell2016}. While the EMM is linearly correlated with the three physical factors in Table \ref{tab:metrics}, it does not scale linearly with the number or contrast of the interference fringes. A large EMM value, indicative of equal mixing between channels, may lead to blurred or obfuscated patterns while a moderate EMM value may show richer interchannel interference fringes. The visibility and structure of these patterns depend heavily on how the contributing events from each channel overlap in the momentum plane. The physical factors help quantify this overlap, while their correlation coefficients indicate their relevance to the \textit{observed} interference. EMM, used alongside these factors, predicts whether interchannel interference is likely to occur, but not how pronounced it will be.
EMM also helps assess whether comparable channels are necessary to observe rich interference i.e., multiple distinct patterns with strong fringe contrast. In this section, we evaluate how well EMM and its correlations explain the observed interference structures.

To facilitate this discussion, we examine six two-channel PMDs spanning high, moderate, and low EMM values, plus one pair with identical excited-state geometries. High EMM pairs, such as $3d4s$ and $4p5s$, are the most comparable channels and are expected to lead to interference patterns that span the momentum region occupied by events of both constituent channels. If the channels occupy similar regions, one may see blurred or superimposed channel interference fringes.
Low EMM pairs, such as $3p4d$, are dominated by a single channel and typically yield sharper fringes confined to the momentum region of the dominant channel, resembling single-channel PMDs. Moderate EMM pair such as $3p3d$, $4d5s$ and $4s5s$ show slight dominance by one channel, making fringe contrast and spatial distribution harder to predict. Where momentum regions do not overlap, we expect crisper fringes; elsewhere, interference may appear blurrier. For the $4s5s$ pair, which occupies typical $s$-state regions along the axes and near the origin, channel-only interference is expected primarily along these regions.
Importantly, all PMDs exhibit intrachannel interference to varying degrees. In cases with high or moderate EMM, this may be partially obscured by overlapping interchannel interference.

Fig.~\ref{fig:coherent} shows the fully coherent two-channel PMD for the six two-channel sums of focus (see bold-text channels in Table \ref{tab:metrics}). The interference comprises of a combination of all aforementioned interference types, and the coherent PMDs exhibit additional substructure in comparison to their counterparts in Fig.~\ref{fig:incoherent}. To discern the interference pattern more clearly, we compute the parameter $\mathcal{P}_{(ccc)}-\mathcal{P}_{(iii)}$, a difference of the fully coherent and incoherent sums, displayed in Fig.~\ref{fig:difference}. This figure serves as a reference for gauging the contributions of each interference type across momentum regions and establishing their overall hierarchy.

The influence of channel intensity on EMM and consequently on observable interference is evident when comparing the PMDs with $3p3d$ [Fig.~\ref{fig:difference}(a)] and $3d4s$ [Fig.~\ref{fig:difference}(d)]. Both resemble the single-channel $3d$ distribution [Fig.~\ref{fig:singlechannel}(a)], yet the $3d4s$ PMD shows a more balanced contribution due to equal channel intensity $M_\text{diff}$, with maxima not only in the $3d$-dominated regions but also at the origin from the $4s$-state. Although the overall interference patterns are similar, faint checkerboard structures in the second and fourth quadrants of panel (a) appear slightly blurred in panel (d), consistent with stronger interchannel interference. 

A similar contrast arises between $3p4d$ [Fig.~\ref{fig:difference}(b)] and $4d5s$ [Fig.~\ref{fig:difference}(c)], both involving $d$-states but differing in fringe clarity: $3p4d$ exhibits crisp spine lines skewed by hyperbolic fringes, reflecting its low EMM and large intensity contrast, while $4d5s$ shows similar but less pronounced features due to moderate EMM and slight  $4d$-dominance. Likewise, the PMDs of $4s5s$ [Fig.~\ref{fig:difference}(e)] and $4p5s$ [Fig.~\ref{fig:difference}(f)] share $5s$-state characteristics, but $4p5s$ displays more complex fringe structures particularly in the first quadrant where additional interference contributions smear the pattern, consistent with its higher EMM.

These observations raise broader questions: How do distinct interference types manifest in the parallel momentum plane e.g. which regions they occupy, the patterns they produce, and how the building blocks interact with each other? Which interference types are most sensitive to the factors investigated; for instance, can smaller energy differences between channels enhance or suppress specific interference features? Systematic analysis of these trends may yield criteria for predicting interference visibility, such as the conditions under which certain patterns dominate or become obscured. 

Consequently, we will now systematically investigate the four interference types. For each, we discuss the high, moderate, and low EMM cases, that is, two-channel PMDs where both channels \textit{appear} to contribute fairly equally (expected largest channel interference), where one channel \textit{slightly} dominates the other (difficult to predict channel interference and clarity of fringes) and where one channel clearly appears to dominate over the other (expect little interchannel interference and otherwise crisp fringes). We note any trends in interference, such as the number of phase variations and contrast of the fringes, as the EMM value changes. The EMM value provides a roadmap by which we can systematically examine the extent to which the three physical factors play a role. 

\subsubsection{Channel-Only Interference}
\begin{figure}
    \centering
    \includegraphics[width=\columnwidth]{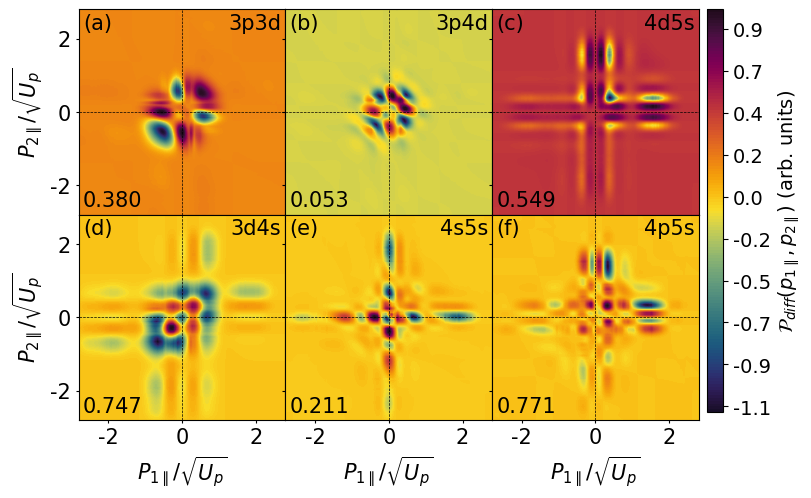}
    \caption{Difference between the RESI distribution $\mathcal{P}_{iic} (p_{1\parallel}, p_{2\parallel})$ with incoherent symmetrization and incoherent summation of events but with channels added coherently [see Table~\ref{tab:possiblesums}(d)], and the fully incoherent probability density $\mathcal{P}_{iii} (p_{1\parallel}, p_{2\parallel})$ for select pairs of channels, computed with the same driving field parameters as in Fig.~\ref{fig:pulseshape}. The excited states for each pair of channels is shown in the top-right corner. The axes are indicated with black dashed lines. The equal mix metric value is indicated in the bottom-left corner of each panel. The signal in each panel has been normalized with regard to its maximum. }
    \label{fig:channel}
\end{figure}

Channel-only interference, depicted schematically in Fig.~\ref{fig:schematic2}(a) arises from phase differences such as $\alpha_{l,l}^{(C_n,C_m)}$ [Eq.~\eqref{eq:channelonly}]. This consists of building blocks which will mainly lead to linear fringes parallel to the axes and potentially along the diagonals with various phase shifts depending on the amount of overlap of the contributing channels' events in the momentum plane. The interference patterns for the six channel pairs in focus are displayed in Fig.~\ref{fig:channel}.

The PMDs of high EMM cases such as  $3d4s$ and $4p5s$ [Fig.~\ref{fig:channel}(d), (f) respectively] exhibit clear signatures from both constituent channels as expected, along with numerous phase changes. Note that since EMM is not exactly 1 for any case, one channel always slightly prevails over the other. This is best exemplified by the $3p3d$ and $3d4s$ cases [Fig.~\ref{fig:channel}(a), (d)], where the PMDs are primarily shaped by the $3d$ state. With $3d4s$, the $4s$ contribution subtly distorts the distribution in line with its slightly higher EMM value. These similarities suggest comparable interference patterns. These phase changes arise from constructive and destructive interference between channels, driven by the building blocks that induce numerical shifts. A higher number of phase changes implies strong channel interference. The $4p5s$ [Fig.~\ref{fig:channel}(f)] case has a larger overlap parameter than $3d4s$ [Fig.~\ref{fig:channel}(d)], leading to a greater number of phase variations.

Similar phase changes occur in almost all panels of Fig.~\ref{fig:channel}. The $4d5s$ case [Fig.~\ref{fig:channel}(c)], with moderate EMM and lower overlap than $4p5s$, shows fewer phase variations and is dominated by the $4d$ state due to the large $M_\text{diff}$ value, associated with the relative channel intensity. Consequently, phase shifts occur mainly in the first quadrant, parallel to the axes, rather than directly along them as typical for an $s$-state. The $3p3d$ distribution [Fig.~\ref{fig:channel}(a)], also with moderate EMM, is shaped predominantly by the $3d$ channel. Since both $3p$ and $3d$ PMDs are concentrated near the origin, the channel interference is also localized in this region. Phase variations are minimal since the events from both channels do not overlap much when mapped to the parallel momentum plane.

In contrast, the $4s5s$ case [Fig.~\ref{fig:channel}(e)], with a yet lower EMM and strong overlap between $s$-state distributions, exhibits phase variations along the axes. Finally, the $3p4d$ case [Fig.~\ref{fig:channel}(b)], despite its lowest EMM, exhibits numerous phase differences comparable to $4s5s$. Here, the large energy difference enhances the $\Delta \tau_{(n,m)}$ term and amplifies the influence of the building blocks, including the $\alpha^{(\text{shift})}$ contribution. Due to the large $M_\text{diff}$ value, the PMD is dominated by the $4d$ state and concentrated in its characteristic momentum region [see Fig.~\ref{fig:singlechannel}(a) in the Appendix for the single-channel $4d$ PMD]. Strong spatial overlap between $3p$ and $4d$ events makes the interference features more prominent in direct contrast with Fig.~\ref{fig:singlechannel}(a), where such overlap is minimal.

\subsubsection{Channel-Exchange Interference}

\begin{figure}
    \centering
    \includegraphics[width=\columnwidth]{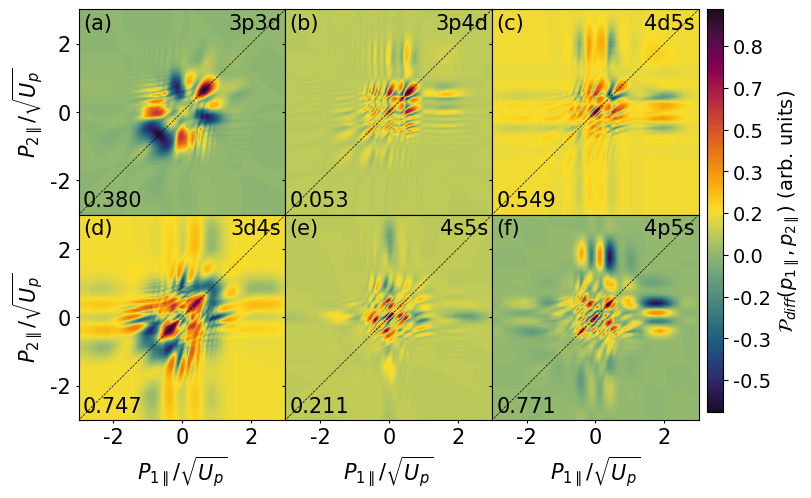}
    \caption{Difference between the RESI distribution $\mathcal{P}_{cic} (p_{1\parallel}, p_{2\parallel})$ (coherent sum over channels and symmetrization, incoherent sum over events [see Table ~\ref{tab:possiblesums}(b)]), and the fully incoherent probability density $\mathcal{P}_{iii} (p_{1\parallel}, p_{2\parallel})$ for select pairs of channels, computed with the same driving field parameters as in Fig.~\ref{fig:pulseshape}. The excited states for each pair of channels is shown in the top-right corner. The diagonal $p_{1\parallel}=p_{2\parallel}$ is indicated with a dashed black line. The equal mix metric value is indicated in the bottom-left corner of each panel. The signal in each panel has been normalized with regard to its maximum.}
    \label{fig:channelexchange}
\end{figure}

The second type of interference is channel-exchange, shown in Fig.~\ref{fig:schematic2}(b).
This interference type is the dominant contributor to the fully coherent PMDs - many of the exchange features displayed in Fig.~\ref{fig:channelexchange} are prominent in Fig.~\ref{fig:difference}. Across all cases, the PMDs exhibit diagonal fringes skewed by hyperbolae, consistent with the building blocks associated with channel-exchange [Eq.~\eqref{eq:channelexchange}]. These arise particularly from the channel-independent $\alpha^{(\text{exch})}$ term and the $\alpha^{p_1 \leftrightarrow p_2}$ term, which contributes strongly regardless of $\Delta t_{(n,m)}$; consequently, this building block produces such patterns for all pairwise channel sums. Eq.~\eqref{eq:channelexchange} additionally predicts linear and `feathery` fringes that may overlap to form checkerboard-like structures accompanied by possible phase jumps. 

The dependence of these features on the effective mixing magnitude (EMM) is evident across cases. For low EMM, such as $3p4d$ [Fig.~\ref{fig:channelexchange}(b)], the PMD is dominated by the $4d$ state and closely resembles the single-channel result [see Fig.~\ref{fig:appendix3}(c) compared to Fig.~\ref{fig:appendix3}(a)]. The fringes are crisp with minimal phase fluctuations, as the large magnitude difference suppresses channel interference, leaving only the channel-independent exchange contribution from $\alpha^{(\text{exch})}_{(\mathbf{p}_1,\mathbf{p}_2)}$.

In contrast, the moderate EMM case $4d5s$ [Fig.~\ref{fig:channelexchange}(c)] exhibits similar closely-spaced diagonal fringes but with increased waviness and phase jumps, particularly at larger momenta, and reduced contrast in the central fringe. The central spine, a signature of intrachannel exchange interference, becomes attenuated with stronger interchannel mixing. The diagonal fringes of the other moderate case, $4s5s$ [Fig.~\ref{fig:channelexchange}(e)] resemble those of the high EMM $4p5s$ case [Fig.~\ref{fig:channelexchange}(f)], although there is blurring in quadrant one of the latter arising from events of the $4p$-state being mapped to larger momentum regions. In both, phase variations blur the fringes and diminish the prominence of the central spine.

Finally, the $3p3d$ and $3d4s$ cases [Figs.~\ref{fig:channelexchange}(a,d)] exhibit similar interference structures since they are both dominated by the $3d$ state to varying degrees. The former case shows pronounced diagonal fringes and a strong spine-like feature, while the latter exhibits a broader spine spanning quadrants one and three, modified by the $4s$ contribution. This is expected given $3d4s$ has a higher EMM than $3p3d$. Both also feature feathery, fishbone-like fringes in the second and fourth quadrants, arising from the $\alpha^{(\mathbf{p}_1\leftrightarrow \mathbf{p}_2)}{\Delta \tau{(n,m)}}$ term whose effect is washed out to varying degrees in other panels except for $3d4s$ where they overlap faintly to form a checkerboard pattern. These fishbone structures are linked to the temporal component in $\alpha^{(A^2)}$ phase shift [Eq.~\ref{eq:alphaA2}], associated with the delay $\Delta \tau{(C_n,C_m)}$ between channels.

\subsubsection{Channel-Temporal Interference}

\begin{figure}
    \centering
    \includegraphics[width=\columnwidth]{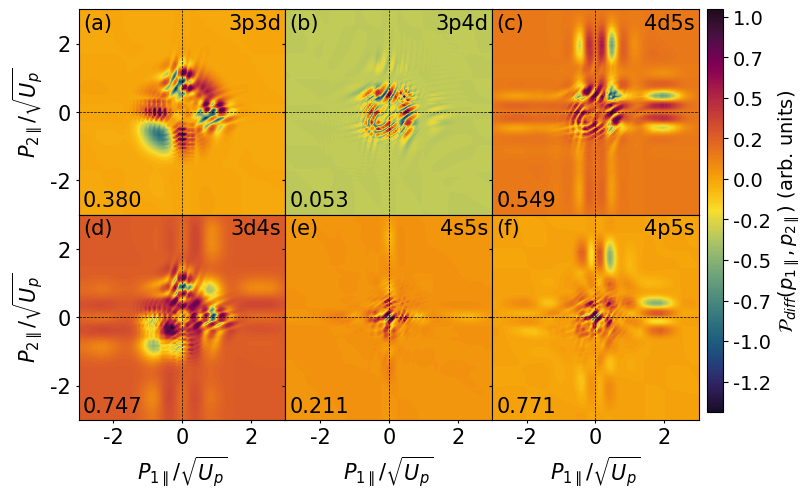}
    \caption{Difference between the RESI distribution $\mathcal{P}_{icc} (p_{1\parallel}, p_{2\parallel})$ (coherent sum over channels and events, unsymmetrized [see Table ~\ref{tab:possiblesums}(c)]), and the fully incoherent probability density $\mathcal{P}_{iii} (p_{1\parallel}, p_{2\parallel})$ for select pairs of channels, computed with the same driving field parameters as in Fig.~\ref{fig:pulseshape}. The excited states for each pair of channels is shown in the top-right corner. The axes are indicated with a dashed black line. The equal mix metric value is indicated in the bottom-left corner of each panel. The signal in each panel has been normalized with regard to its maximum.}
    \label{fig:channeltemporal}
\end{figure}

Channel–temporal interference, depicted schematically in Fig.~\ref{fig:schematic2}(c) and described by phase differences such as $\alpha_{l,r}^{(\mathcal{C}_n, \mathcal{C}_m)}$ [Eq.~\eqref{eq:channeltemporal}], comprise of building blocks which predominantly influence the total interference at large momenta - compare temporal patterns displayed in Fig.~\ref{fig:channeltemporal} to the total interference shown in Fig.~\ref{fig:difference}. The building blocks are expected to lead to shapes such as circular or oval fringes, linear fringes parallel to the axes and possible checkerboard-like structures. The degree and nature of channel–temporal interference varies strongly between cases.

The low-EMM case $3p4d$ [Fig.~\ref{fig:channeltemporal}(b)], characterized by a large $M_\text{diff}$ value, unsurprisingly closely resembles the single-channel $4d$ temporal interference pattern [see Fig.~\ref{fig:singletemporal}(c) in the Appendix].A similar temporal interference pattern coming from the dominance of the $4d$- state occurs for $4d5s$ [Fig.~\ref{fig:channeltemporal}(c)], which also features distinctive oval fringes in the third quadrant. In contrast, the moderate-EMM $4s5s$ and high-EMM $4p5s$ cases [Figs.~\ref{fig:channeltemporal}(e),(f)] exhibit only weak, low-contrast fringes near the origin, reflecting poor momentum-space overlap between temporally-displaced events [see Fig.~\ref{fig:singletemporal}(e),(f) and refer to Figs.~15(g,h) in \cite{Hashim2024} for a more in-depth discussion]. The richest temporal interference occurs for $3p3d$ and $3d4s$ [Figs.~\ref{fig:channeltemporal}(a),(d)], both involving the $3d$ state. The interference contrast is stronger for the $3d4s$ distribution than the $3p3d$ and blurrier given the temporal shifts arising from $\Delta \tau_{(C_m,C_n)}$. Both these PMDs contain diverse structures: linear and diagonal fringes, multiple phase jumps as well as overlaid patterns from both channels. For example, the $3p3d$ PMD exhibits temporal interference characteristic of the $3d$ state [see Fig.~\ref{fig:singletemporal}(b)], as well the similar pattern from the $3p$ state [see Fig.~\ref{fig:singletemporal}(a)] which is overlaid and slightly out of phase.

Overall, interchannel temporal interference introduces greater blurring and complexity than intrachannel interference because of the larger total time displacement $\Delta \tau_{\text{(total)}}$, which amplifies certain phase-building blocks, such as the circular substructure from the $\alpha^{(\text{ene})}$ term which now manifests even for lower momenta as in Fig.~\ref{fig:channeltemporal}(c). 

\subsubsection{Channel Exchange-Temporal (Combined) Interference}
\begin{figure}
    \centering
    \includegraphics[width=\columnwidth]{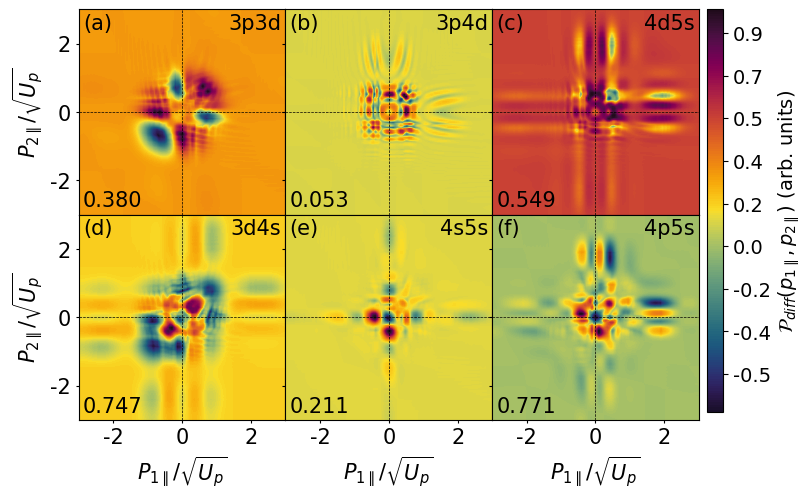}
    \caption{Difference between the two-channel RESI distribution $\mathcal{P}_{ccc, \varepsilon\varepsilon'} (p_{1\parallel}, p_{2\parallel})$ [Eq.~\eqref{eq:combinedP}] which involves summing an event in the first channel with the symmetrized counterpart of a time-delayed event in another channel, and the fully incoherent probability density $\mathcal{P}_{iii} (p_{1\parallel}, p_{2\parallel})$ for select pairs of channels, computed with the same driving field parameters as in Fig.~\ref{fig:pulseshape}. The events chosen are $p_3o_4$ and $p_4o_5$ so $\Delta \tau_{(\varepsilon, \varepsilon')} = \pi/\omega$. The excited states for each pair of channels is shown in the top-right corner. The axes are indicated with a dashed black line. The equal mix metric value is indicated in the bottom-left corner of each panel. The signal in each panel has been normalized with regard to its maximum.}
    \label{fig:channelcombined}
\end{figure}

Channel-combined interference, schematically shown in Fig.~\ref{fig:schematic2}(d) and arising from phase differences such as $\alpha_{l,u}^{(\mathcal{C}_n,\mathcal{C}_m)}$ [Eq.~\eqref{eq:channelcombined}], produces very faint patterns and contributes least to the overall interference. Comparison of Fig.~\ref{fig:channelcombined}, showing the channel-combined interference for the studied pairs, with the channel-only interference in Fig.~\ref{fig:channel}, reveals that the overall PMD shapes and locations remain nearly unchanged. Most visible structures originate from channel interference, while the combined component adds only fine fringes and slight blurring. The largest deviation occurs for the lowest EMM case, $3p4d$ [Fig.~\ref{fig:channelcombined}(b)], whose PMD closely matches the $4d$ interchannel interference pattern [see Fig.~\ref{fig:appendix3}(d)] at smaller momenta skewed by the $3p$ interchannel interference [see Fig.~\ref{fig:appendix3}(b)] at slightly larger momenta, causing numerous fine fringes. It is also the only case displaying the “wing” structures reported in \cite{Hashim2024}, which are suppressed in other panels by stronger channel–temporal interference patterns. The wings are pure intrachannel features which are not obfuscated due to the lack of overlap of events in momentum space from these channels, and because combined-channel interference does not depend on the $\alpha^{\text{shift}}$ building block unlike other interference types.

In the remaining cases, the main difference between the channel-only and channel-combined PMDs is a phase inversion caused by the additional time delay $\Delta \tau_\varepsilon$, which alters conditions for constructive and destructive interference in some building blocks of Eq.~\eqref{eq:channelcombined}. Fine diagonal fringes also appear, with gradients and offsets determined by the underlying phase-building blocks (see \cite{Hashim2025} for details on fringe spacing). The clearest example is the high-EMM case $3d4s$ [Fig.~\ref{fig:channelcombined}(d)], which exhibits shallow diagonal fringes along the positive momentum half-axes. Other high and moderate-EMM cases, $4p5s$, $4s5s$, and $4d5s$ [Figs.~\ref{fig:channelcombined}(f),(e),(c)], show only faint low-contrast fringes near the origin, consistent with limited overlap between temporally displaced channel events. The moderate-EMM $3p3d$ case [Fig.~\ref{fig:channelcombined}(a)] displays fine fishbone-like fringes in quadrants one and three, arising from partial temporal overlap of temporally-displaced and symmetrized events in the contributing channels.

\section{Conclusions}
\label{sec:conclusions}

In this paper, we investigate multichannel interference in recollision excitation with subsequent ionization (RESI) using statistical measures to determine the extent to which different excitation pathways contribute to the total PMDs and examine some of the factors influencing their shape and intensity. We also derive analytical interchannel interference conditions, extending our previous work focused on the intrachannel phase differences \cite{Hashim2024,Hashim2025} to an inter-channel setting. As an application, we discuss two-channel interference in Argon with a linearly polarised few-cycle pulse, although the method is general and can be extended to arbitrary fields and interfering channels. 
Due to the overwhelming number of processes involved, employing statistical measures constitutes a more effective strategy than purely analytical arguments, as they determine whether inter-channel interference is appreciable or physically relevant. 

In previous work \cite{Maxwell2015,Maxwell2016}, two-channel sums were investigated employing ad-hoc arguments, which led to the empirical observation that the spacing of the inter-channel fringes is inversely proportional to the energy difference between two excitation channels. Here,  we approach the problem from a more systematic angle and show that, for inter-channel interference to be significant, three other conditions must be fulfilled: (1) the relative intensities of the contributing channels' PMDs must be comparable; (2) there must be a large overlap [Eq.~\eqref{eq:overlap}] between the excited states' radial wavefunctions and (3) there must be a large overlap between the mapping of the dominant events onto the $p_{1\parallel}p_{2\parallel}$ plane from the contributing channels.

This conclusion is reached by defining an `Equal Mix Metric' which quantifies the extent to which a two-channel PMD is an equal mix of the contributing channels, or whether one channel dominates, and then computing the correlation with the energy difference, overlap and relative intensities. The third condition is determined from unexpected behavior of the two-channel PMD interference. The EMM is based on the Earth Mover's Distance, a statistical measure used to quantify the dissimilarity between two probability distributions. These are commonly used in several areas of knowledge, such as computer vision \cite{Zhang2020emd}, biology \cite{Orlova2016}, ecology \cite{Kranstauber2017}, chemistry \cite{Wang2023emd} and particle physics \cite{Komiske2019}, but not in strong-field and attosecond physics. Nonetheless, it is potentially a valuable tool for understanding to which extent different quantum 
pathways contribute to the underlying dynamics in an interferometric process.

Interchannel interference is governed primarily by target properties rather than field symmetries and is challenging to disentangle due to the vast number of contributing pathways. Nevertheless, following the methodology of Refs.~\cite{Hashim2024,Hashim2025}, we derive analytic interference conditions for channel-only, channel-exchange, channel-temporal, and combined (temporal–exchange) interferences. This extends the classification of two-electron quantum-interference mechanisms in RESI within the SFA to arbitrary driving fields, completing the systematic framework introduced in Refs.~\cite{Hashim2024,Hashim2025}. The resulting interference patterns exhibit familiar building blocks from the intrachannel case such as spine, hyperbolic, and fishbone structures but when multiple excitation channels are involved these patterns can overlap, distort, and blur, rendering a fully analytic characterization difficult. Among the various types, channel-exchange interference is dominant, while the remaining interference types play a secondary role, consistent with the trends observed for intrachannel interference~\cite{Hashim2024}.

The toolbox of interchannel interference conditions derived the present work is general. It is applicable to arbitrary fields, and may provide guidelines for constructing two-electron interferometric schemes, similar to the configurations employed in one-electron pump-probe schemes. This may become a powerful tool, as preliminary results show that inter-channel interference has potential for reshaping correlated electron-momentum distributions \cite{Hao2014,Maxwell2016}. If two excitation channels can be pre-selected according to well-defined criteria, this may inform the construction of bound-state coherent superpositions in RESI. The resulting interference patterns are also influenced by the characteristics of the driving field. Therefore, our inter-channel results connect target structure, field structure and quantum phases in a unified way. Consequently, the choice of preparation scheme or the mechanism used to induce specific transitions may substantially alter the interference outcome. Thus, these insights may serve as a foundation for tailoring multi-electron quantum interference. Moreover, in a realistic scenario, interference will only take place for quantum pathways resulting in the same final compound ionic-electronic state. Therefore, for some pathways recombination to ensure this may need to be incorporated in the present model.  

Finally, this work more broadly illustrates how techniques originating from statistical and computational physics can be leveraged to extract physical insight caused by microscopic dynamics from correlated systems by comparing predicted and measured distributions.
This is critical for applications such as quantum-enhanced imaging, where pattern comparison is critical. Quantum-boosted interferometric setups exist in areas of knowledge as diverse as gravitational wave detection
\cite{Aasi2013} and nonlinear microscopy \cite{Casacio2021}. With the advent of strong-field physics with quantum light, harnessing this advantage is expected to become important in extreme quantum optics as well, although the
range of applications is broader. Thus, it will be desirable to have tools such as the EMD to detect changes in probability distributions. Although most studies of intense laser-matter interaction with quantum light have been done for high-order harmonic generation \cite{Rasputnyi2024,Lewenstein2020} and above threshold ionization \cite{Fang2023, Wang2023, RiveraDean2025}, NSDI computations with bright squeezed vacuum do exist \cite{Liu2025}. 
NSDI and more specifically, RESI serves as a minimal yet non-trivial process for exploring quantum electron-electron correlations in strong laser fields. Thus such methods may be particularly of interest for attosecond quantum technologies. There is currently a major interdisciplinary effort to unite quantum technologies and attosecond physics. So far, most studies of quantum correlation have looked at electron-field (see e.g. \cite{Bhattacharya2023, RiveraDean2022, Stammer2023} amongst others) or electron-ion \cite{Vrakking2021, Vrakking2022, Ivanov2024, Majorosi2017, Ruberti2024} entanglement, and there is large untapped potential regarding electron-electron entanglement, with only a few recent studies \cite{Shobeiry2024, Younis2024,Maxwell2022, Zhang2025} The systematic analysis of interference in conjunction with the statistical tools used provide a powerful way to probe and even tailor such correlations.

\acknowledgements
Discussions with Ian Ford, Bradley Augstein and Rebecca Tenney are gratefully acknowledged. This work was partly funded by Grant No. EP/T019530/1, from the UK Engineering and Physical Sciences Research Council (EPSRC) and by UCL.
\sloppy
%\bibliography{ref,Maciej,OAMinSF}

\appendix*

\subsection{Single-channel PMDs}

\begin{figure}[!h]
    \includegraphics[width=\columnwidth]{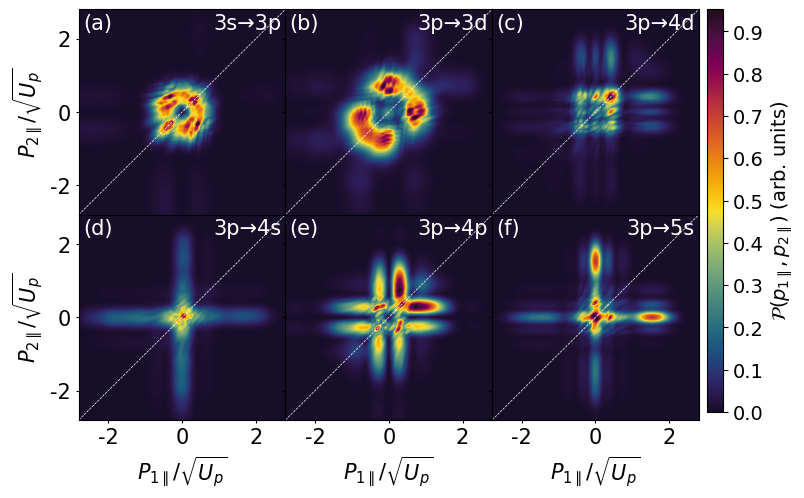}
    \caption{Fully coherent single-channel momentum distributions $\mathcal{P}_{cc}(p_{\parallel}, p_{2\parallel})$ for all channels in Table~\ref{tab:channels}. The transitions are indicated in the top-right corner. The diagonal is indicated by the white dashed line. All other parameters are the same as in Fig.~\ref{fig:pulseshape}. }
    \label{fig:singlechannel}
\end{figure}

\begin{figure}[!h]
    \includegraphics[width=\columnwidth]{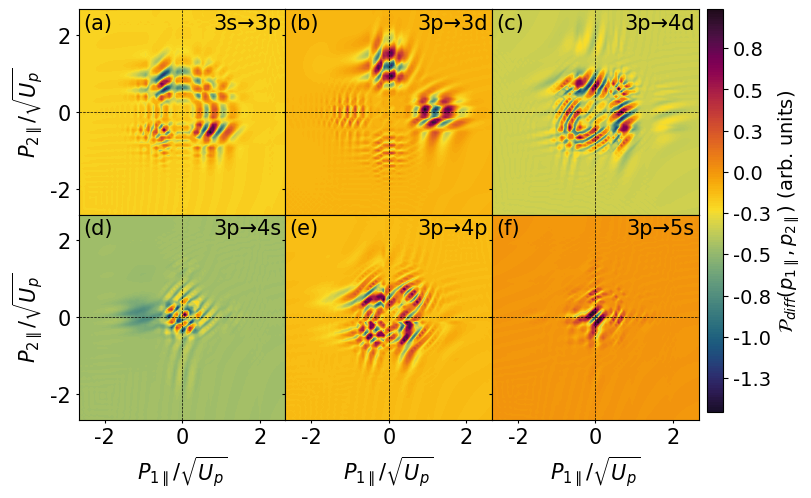}
    \caption{Difference between the RESI distributions $\mathcal{P}_{ic}(p_{\parallel}, p_{2\parallel})$ where events are summed over coherently, and symmetrization is incoherent, and the fully incoherent sum $\mathcal{P}_{ii}(p_{\parallel}, p_{2\parallel})$, for each of the six channels in Table~\ref{tab:channels}. Each panel is normalized with regard to its maximum. The transitions are indicated in the top-right corner. The axes are indicated with black dashed lines. All other parameters are the same as in Fig.~\ref{fig:pulseshape}.}
    \label{fig:singletemporal}
\end{figure}

\begin{figure}[h]
    \includegraphics[width=\columnwidth]{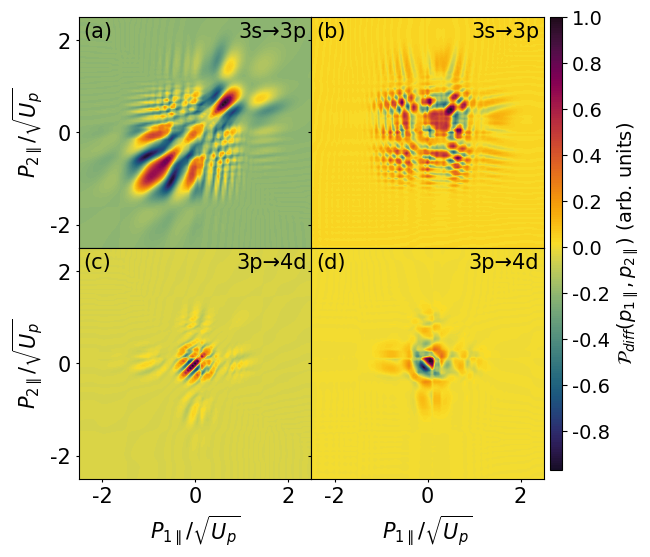}
    \caption{RESI distributions $\mathcal{P}_{ci}(p_{\parallel}, p_{2\parallel})$ - $\mathcal{P}_{ii}(p_{\parallel}, p_{2\parallel})$ isolating intrachannel exchange interference [panels (a), (c)], and $\mathcal{P}_{cc, \varepsilon\varepsilon'}(p_{\parallel}, p_{2\parallel})$ - $\mathcal{P}_{ii}(p_{\parallel}, p_{2\parallel})$ isolating intrachannel combined interference taking the two most dominant events for the pulse in question $p_3o_4$, $p_4o_5$ [panels (b), (d)], for the $3s\rightarrow3p$ [panels (a), (b)] and the $3p\rightarrow4d$ [panels (c), (d)] transitions. Each panel is normalized with regard to its maximum. The transitions are indicated in the top-right corner. The axes are indicated with black dashed lines. All other parameters are the same as in Fig.~\ref{fig:pulseshape}.}
    \label{fig:appendix3}
\end{figure}

Here we provide the single channel PMDs for the six channels listed in Table \ref{tab:channels}, used to compute the two-channel sums in this paper. These were originally presented in \cite{Hashim2024} and are reproduced here in the same scale, and colormap for ease of reference. 
Fig.~\ref{fig:singlechannel} displays the fully coherent PMDs while Fig.~\ref{fig:singletemporal} displays the temporal interference.
Finally, we give the intrachannel exchange and combined interference plots in Fig.~\ref{fig:appendix3} for the $3s\rightarrow3p$  and $3p\rightarrow4d$ channels. This supplements the discussion of the $3p4d$ two-channel interference. 

\subsection{Earth Mover's Distance}
\label{Sec:EMDappendix}
To quantify similarity between the photoelectron momentum distributions across the channels, pairwise Earth Movers Distances (EMDs) are computed. The EMD is mathematically equivalent to and otherwise known in literature as the Wasserstein-1 $(W_1)$ distance \cite{Givens1984}. The signal for each channel $\mathcal{C}$ is represented by a non-negative 2D matrix defined on a common momentum grid $(p_\parallel, p_\perp)$. Prior to the distance computation, we applied the following preprocessing steps. First, each matrix is symmetrized with respect to the anti-diagonal by averaging the original matrix with its anti-diagonal reflection. This enforces the expected (and qualitatively observed) symmetry of the underlying signal and reduces any noise-induced asymmetries arising for example from the resolution. Second, the matrices are downsampled by block averaging to reduce computational cost while preserving large-scale spatial structure. Finally, each matrix is normalized such that it defines a discrete probability distribution 
\begin{equation}
\mu^{(\mathcal{C})} = \sum_{i=1}^{N} \sum_{j=1}^{M} \mu_{p_{\parallel i}p_{\perp j}}^{(\mathcal{C})} = 1,
\end{equation}
where the grid is identified by the set of points $\{(p_{\parallel i},p_{\perp j})\}_{i=1,...,N;j=1,...,M} \subset \mathbb{R}^2$. This is required for optimal transport computations \cite{Rubner2000}. Each grid cell is treated as a point in 2D Euclidean space. The ground cost between two momentum-space bins $(p_{\parallel i}, p_{\perp j})$ and $(p_{\parallel i'}, p_{\perp j'})$ is defined as their Euclidean distance
\begin{widetext}
\begin{equation}
    c[(p_{\parallel i}, p_{\perp j}),(p_{\parallel i'}, p_{\perp j'})]=  \sqrt{(p_{\parallel i}- p_{\parallel i'})^2 +(p_{\perp j}- p_{\perp j'})^2}.
\end{equation}

For each pair of channels $\mathcal{C}_n$ and $\mathcal{C}_m$, the EMD is computed between the corresponding normalized distributions using this precomputed ground cost matrix. The EMD quantifies the minimum cost required to transform one spatial distribution to another and is computed as
\begin{equation}
W_1\!\left(\mu^{(\mathcal{C}_n)}, \mu^{(\mathcal{C}_m)}\right) = 
\min_{\gamma \in \Gamma\left(\mu^{(\mathcal{C}_m)}, \mu^{(\mathcal{C}_n)}\right)}
\sum_{(p_{\parallel i,}, p_{\perp j}), (p_{\parallel i'}, p_{\perp j'})} 
\gamma_{(p_{\parallel i}, p_{\perp j}),(p_{\parallel i'}, p_{\perp j'})}
\, c\!\left[(p_{\parallel i}, p_{\perp j}),(p_{\parallel i'}, p_{\perp j'})\right]
    \label{eq:wasserstein1}
\end{equation}
\end{widetext}
where $\gamma$ denotes a `transport plan' specifying how the probability mass $\mu_{p_{\parallel i}, p_{\perp j}}$, is redistributed between discrete momentum-space bins. The set $\Gamma (\mu^{(\mathcal{C}_n)}, \mu^{(\mathcal{C}_m)})$ contains all transport plans that exactly reproduce the source and target distributions i.e. the total mass leaving each bin of the source and arriving at each bin of the target is fixed.
The resulting pairwise distances form a symmetric matrix. This yields a metric that captures differences in the spatial structure of the PMD signal between channels (as opposed to simple pointwise differences in intensity) and is therefore sensitive to collective shifts of the PMD signal. 

\subsection{Pearson Correlation Coefficient}
\label{Sec:pccAppendix}
The Pearson correlation coefficient (PCC) is computed to investigate any correlations between the EMD values and the other factors expected to influence interference such as the energy difference. 
Generically, for any two variables $X$ and $Y$ with paired entries $(X_i, Y_i)$ and means $\bar{X}, \bar{Y}$, the PCC is defined as
\begin{equation}
r_{XY} = \frac{\sum_i (X_i - \bar{X})(Y_i - \bar{Y})}
{\sqrt{\sum_i (X_i - \bar{X})^2}\;\sqrt{\sum_i (Y_i - \bar{Y})^2}},
\label{eq:PCC}
\end{equation}

This results in a scalar value $r_{XY} \in [-1,1] $ quantifying the strength of the linear correlation between the variables X and Y.

\end{document}